\documentclass[showpacs, twocolumn, preprintnumbers,amsmath,amssymb]{revtex4}
\usepackage[utf8]{inputenc}
\usepackage{mathrsfs}
\usepackage{graphicx}
\usepackage{subfigure}
\usepackage{indentfirst}
\usepackage{amsfonts}
\usepackage{amssymb}%
\usepackage{color}
\usepackage{slashed}
\usepackage{extarrows}

\def\al{\alpha}
\def\be{\beta}
\def\ga{\gamma}
\def\de{\delta}

\def\la{\lambda}

\def\si{\sigma}

\def\ps{\psi}

\def\Ga{\Gamma}

\def\Si{\Sigma}

\def\Ps{\Psi}

\def\mn{{\mu\nu}}

\def\prt{\partial}

\def\lsim{\mathrel{\rlap{\lower4pt\hbox{\hskip1pt$\sim$}}
    \raise1pt\hbox{$<$}}}
\def\gsim{\mathrel{\rlap{\lower4pt\hbox{\hskip1pt$\sim$}}
    \raise1pt\hbox{$>$}}}
\def\sqr#1#2{{\vcenter{\vbox{\hrule height.#2pt
         \hbox{\vrule width.#2pt height#1pt \kern#1pt
         \vrule width.#2pt}
         \hrule height.#2pt}}}}
\newcommand{\beq}{\begin{equation}}
\newcommand{\eeq}{\end{equation}}
\newcommand{\bea}{\begin{eqnarray}}
\newcommand{\eea}{\end{eqnarray}}

\newcommand{\nn}{\nonumber\\}

\def\cL{{\cal L}}

\def\lsim{\mathrel{\rlap{\lower4pt\hbox{\hskip1pt$\sim$}}
    \raise1pt\hbox{$<$}}}
\def\gsim{\mathrel{\rlap{\lower4pt\hbox{\hskip1pt$\sim$}}
    \raise1pt\hbox{$>$}}}

\def\prt{\partial}
\def\lrprt#1{\hskip-2pt\stackrel{\leftrightarrow}{\prt}_{#1}\hskip-2pt}

\def\lrdrt#1{\hskip-2pt\stackrel{\leftrightarrow}{D}_{#1}\hskip-2pt}
\def\lrdut#1{\hskip-2pt\stackrel{\leftrightarrow}{D^{#1}}\hskip-2pt}
\def\lrNrt#1{\hskip-2pt\stackrel{\leftrightarrow}{\nabla}_{#1}\hskip-2pt}

\def\mn{{\mu\nu}}

\def\psb{\overline{\ps}{}}

\newcommand{\ie}{{\it i.e.}}
\newcommand{\etc}{{\it etc}}
\newcommand{\eg}{{\it e.g.}}
\newcommand{\etl}{{\it et.al.}}
\newcommand{\hf}{\frac{1}{2}}
\newcommand{\bit}{\begin{itemize}}
\newcommand{\eit}{\end{itemize}}

\providecommand{\Journal}[4] {#1 {\bf#2}, #4 (#3)}
\providecommand{\IJMPA}{Int. J. Mod. Phys. A} %
\providecommand{\CQG}{Class. Quantum Grav.}%
\providecommand{\GRG}{Gen. Rel. Grav.}%
\providecommand{\LRR}{Living Rev. Relativ.}%
\providecommand{\PR}{Phys. Rev.} %
\providecommand{\PRep}{Phys. Rep.} %
\providecommand{\PRL}{Phys. Rev. Lett.} %
\providecommand{\PRD}{Phys. Rev. D.} %
\providecommand{\RMP}{Rev. Mod. Phys.} %
\providecommand{\RPP}{Rept. Prog. Phys.} %
\providecommand{\PLB}{Phys. Lett. B} %
\providecommand{\PZ}{Phys. Z} %
\providecommand{\Pca}{Physica}
\providecommand{\NT}{Nature} %
\providecommand{\AT}{Atoms}
\providecommand{\EPJC}{Eur. Phys. J. C.} %
\providecommand{\JMP}{J. Math. Phys.} %
\providecommand{\JHEP}{JHEP} %
\providecommand{\SPJETP}{Sov. Phys. JETP}%
\providecommand{\IJMPA}{Int. J. Mod. Phys. A}%
\begin{document}

\title{\Large {The analogy of the Lorentz-violating fermion-gravity and fermion photon couplings}}

\author{Cheng Ye}
\affiliation{Department of Mathematics and Physics, North China Electric Power University, Beijing 102206, China}
\author{Zhi Xiao}
\email{spacecraft@pku.edu.cn}
\affiliation{Department of Mathematics and Physics, North China Electric Power University, Beijing 102206, China}

\begin{abstract}
By adopting a methodology proposed by R.J. Adler \etl, we study the interesting analogy between the fermion-gravity and the fermion-electromagnetic interactions in the presence of the minimal Lorentz-violating (LV) fermion coefficients. The one-fermion matrix elements of gravitational interaction (OMEGI) are obtained with a prescribed Lense-Thirring (LT) metric assuming test particle assumption. Quite distinct from the extensively studied linear gravitational potential, the LT metric is an essentially curved metric, and thus reveals the anomalous LV matter-gravity couplings as a manifestation of the so-called gravito-magnetic effects, which go beyond the conventional equivalence principle predictions.
By collecting all the spin-dependent operators from the OMEGI with some reasonable assumptions, we get a LV non-relativistic Hamiltonian,
from which we derive the anomalous spin precession and gravitational acceleration due to LV.
Combined these results with certain spin gravity experiments, we get some rough bounds on several LV
parameters, such as $|3\vec{\tilde{H}}-2\vec{b}|\leq1.46\times10^{-5}\mathrm{eV}$, with some ad hoc assumptions.
\end{abstract}

\maketitle
\section{Introduction}
The classical electrodynamics and its quantum version, QED, are ideal paradigms for modern physics.
As a quantum theory of matter-electromagnetic coupling, QED has reached an unprecedented precision for the
match between theory and observations \cite{Aoyama:2019ryr}.
In fact, the success of QED nourishes many branches of physics, such as the Yang-Mills theory.
Exactly parallel the historical precedent of QED, we expect to gain some insight by
studying matter-gravity couplings in the {\it semi-classical regime} in weak gravity,
given that gravity still resists successful quantization after decades of endeavor.
As a supporting fact, the Einstein field equation and the geodesic equation resemble the Maxwell equation and the Lorentz force law \cite{Mashhoon:2003ax} for a slow moving particle in weak field limit, though this analogy breaks down when gravity is sufficiently strong.
The conceptual reason roots in the peculiar differences between gravity and electromagnetism:
1. {\it gravity is extremely weak and universal}; 2. {\it gravity is highly nonlinear}.

Another motivation for the study lies in the fact that Lorentz violation (LV) may be a testable signal of some unified theory at Planck scale \cite{LVMotivations}.
Many different scenarios lead to LV have been proposed, such as noncommutative field theory \cite{SynderNonCom}, loop gravity \cite{Gambini:1998it}, very special relativity \cite{VSR2006}, etc.
In fact, to systematically study the possible LV effects, an effective field framework incorporating various standard model fields and tiny tensorial coefficients controlling LV has been developed, the Standard-Model Extension, or briefly SME \cite{SMEa}\cite{SMEc}.
This framework facilitates the test of the common foundation of gravity and electroweak theory, namely the Lorentz symmetry.
In SME, the close resemblance between gravity and electromagnetics has been utilized to map a solution
of Maxwell equation with a restricted class of $(k_F)^{\kappa\la\mn}$ term to the solution of the Einstein equation with $\bar{s}_{\mn}$ term \cite{Bailey:2010af},
though the nonlinear acceleration $\vec{a}_\mathrm{NL}$ spoils the exact formal analogy of weak gravity
to electrodynamics even restricted to terms with linear velocity and in the stationary limit.
Combined with the precision measurements of Gravity Probe B \cite{GPB1}\cite{GPB2}, new bounds on $\bar{s}_{\mn}$
have been extracted from the anomalous spin precession caused by the LV gravito-electromagnetic (GEM) fields \cite{LimitS-GPB13}.
With the observation of the structure similarity for the couplings between the gravito-magnetic field
and the LV $\tilde{b}$-type coefficient to intrinsic spin \cite{Tassonspin}, the bounds on $\bar{s}^{0k}-\frac{\al}{m}(\bar{a}^\mathrm{S}_\mathrm{eff})^k$ have been obtained from various comagnetometer experiments \cite{comagnet1}\cite{comagnet2} by reinterpreting $\tilde{b}$ as the gravito-magnetic field caused by the off-diagonal metric perturbation due to LV.

In comparison, in this paper we try to explore the resemblance of the LV fermion-gravity couplings
with the Lense-Thirring metric to the LV fermion-photon couplings
with Lorentz invariant (LI) electromagnetic field in the framework of SME.
In other words, we concentrate on the quantum matter effects induced by LV in this analogy.
For simplicity, we consider only the fermion LV coefficients in the minimal SME and keep the gravity sector intact.
Partially because we are more interested in the LV fermion sector and partially because the LV coefficient $\bar{s}_{\mn}$
in minimal gravity sector can be switched into $\bar{c}_{\mn}$ by a proper field redefinition \cite{TassonMGC},
we do not consider the LV fermion-gravity couplings arising from pure gravity.
No doubt the back reaction of LV matter fields to spacetime geometry necessarily generates LV metric perturbation, and this fact has already been thoroughly explored for the $(\bar{a}_\mathrm{eff})_\mu,~\bar{c}_\mn$ coefficients in \cite{TassonMGC}\cite{Tassonspin}.
However, we make the test particle assumption and ignore the back reaction in our simple setting,
thus no need to worry about the extra modes from diffeomrphism breaking unless the pure gravity sector were also affected by LV.
As for the extra modes due to spontaneous local Lorentz symmetry breaking,
which may play the role of photon or graviton, such as in the bumblebee or cardinal models \cite{SMEc}\cite{BumblebeeGravi}\cite{Bluhm:2004ep}\cite{AKRP2009},
or mediate new forces \cite{NFLV2005},
they suffer severe experimental constraints \cite{dataTable} and lie out of the scope of our present discussion, we disregard them for simplicity.

It is interesting to note that a systematic and thorough treatment of all possible LV matter-gravity couplings, both in formalism and in conceptual issues, have been developed recently \cite{Kost-Backgrounds}, where no room is left for spontaneous local LV with differomorphism invariance.
However, this superficial conflict is because we omit the back reaction of LV matter field to spacetime geometry in the test particle assumption.
Since spontaneous symmetry breaking is assumed, the no-go constraints \cite{SMEc} can also be avoided.
In comparison, the signals beyond-Riemann geometry have been explored with an effective field theory
incorporating all linear fermion-gravity operators up to dimension-5 \cite{Kost-BeyondRieman}, based on the assumption of
local LI but explicit diffeomorphism breaking.
In contrast to Ref.  \cite{Kost-BeyondRieman}, where the typical gravitational acceleration is uniform as
the exploration mainly focus on laboratory experiments on the Earth,
our study assumes the Lense-Thirring metric \cite{LTorigin1},
which is essentially curved and has non-zero source angular momentum.
This setting is particularly suitable for a tentative study of LV gravitomagnetic effects.

As the fermion in the analogy is non-relativistic (NR) for practical purposes,
it seems necessary to perform the Foldy-Wouthuysen (FW) transformation \cite{FWT1950}\cite{Tassonspin}\cite{ZX2018} first, 
however, a different method first proposed in Ref. \cite{Adler2011bg} is adopted,
where the one-fermion matrix elements for a NR fermion scattering off external fields are studied.
The NR feature relies on the assumption that the field quantum carry negligible energy
and fermion quantization is truncated on positive energy states only.
The rest of the paper is arranged as follows.
In Sec. \ref{Intro-GEM}, we briefly review the basic background of gravito-electromagnetism,
an analogy of weak gravity in general relativity to electromagnetism.
In Sec. \ref{LVEMT}, we derive the energy momentum tensor (EMT) for a LV fermion in flat spacetime
as a warm exercise for the discussion of LV matter-gravity couplings,
since gravity couples exactly to the EMT of matter fields, just as photon couples to the electromagnetic current.
In Sec. \ref{NRMEtriC}, we briefly review the formalism describing a LV fermion coupled with gravity
in the weak field approximation.
In Sec. \ref{Analogy}, we outline the main methodology in obtaining the one-fermion matrix elements for a LV fermion coupled with external fields.
To make transparent the analogy, we demonstrate the fermion-photon couplings together with the fermion-gravity couplings.
Possible experimental constraints on LV spin-gravity couplings are discussed in Sec. \ref{ConstrEP},
and we summarize our main results in Sec.\ref{Summ}.

\section{The gravito-electromagnetism}\label{Intro-GEM}
The electromagnetic (EM) analogy for weak gravity can be found in many text books on general relativity \cite{Padmanabhan2010} or review papers \cite{Mashhoon:2003ax}.
The inhomogeneious Maxwell equations and Lorentz force law for a charged particle moving in the EM fields are
\bea\label{MaxwellPE}&&
\nabla\cdot\vec{E}=\frac{\rho_e}{\epsilon_0},\quad
\nabla\times\vec{B}-\frac{1}{c^2}\frac{\prt\vec{E}}{\prt t}=\mu_0\vec{j}_e\\
\label{LorentzF}&&
\frac{d(\ga\vec{v})}{dt}=\frac{e}{m}\left[\vec{E}+\frac{\vec{v}}{c}\times\vec{B}\right].
\eea
For sufficiently weak gravity and slow-moving source, we can expand the metric around Minkowski background
\bea\label{MetricExp}
g_{\mn}=\eta_{\mn}+h_{\mn}.
\eea
when source is stationary $\dot{T}_{\mn}=0$, and in the harmonic gauge $\Ga^\rho\equiv\Ga^\rho_{~\mn}g^{\mn}=0$,
the Einstein field equation $G_{\mn}=\kappa T_{\mn}$ ($\kappa\equiv\frac{8\pi G}{c^4}$)
can be cast into the form similar to (\ref{MaxwellPE}),
\bea\label{GEMa}
\nabla\cdot\vec{E}_g=-\frac{\kappa c^4}{2}\rho_m,\quad
\nabla\times\vec{B}_g-\frac{1}{c}\prt_t\vec{E}_g=-2\kappa c^3\vec{j}_m,
\eea
where $\vec{E}_g\equiv-\nabla\phi_g-\frac{1}{c}\prt_t\vec{A}_g$ is the so-called gravito-electric field,
or just the local gravitational acceleration when $\dot{\vec{A}}_g=0$,
${B_g^i}\equiv{c^2}\epsilon_{ijk}\prt_jh_{0k}$ is the gravito-magnetic field,
and $\rho_m,~\vec{j}_m=\rho_m\vec{v}$ are the matter mass density and mass current, respectively.
It is easy to check that the homogeneous equations similar to $\prt_\mu\tilde{F}^{\mn}=0$ ($\tilde{F}^{\mn}\equiv\hf\epsilon^{\rho\si\mn}F_{\rho\si}$) in electrodynamics are also satisfied,
see Appendix \ref{GraviEM}.
In fact, up to 1PN [$\mathcal{O}(c^{-2})$ for $h_{ij}$], these GEM equations can be further generalized to the case when matter source does have time dependence, as long as the gravitating system is moving slowly, see \cite{EPCWW}\cite{PostNThrone}\cite{Mashhoon:2003ax}.
In that case, one can even derive a formal equation $\left[\nabla^2-\frac{4}{c^2}\prt_t^2\right]\vec{B}_g=0$
for the fields outside the source current,
which may indicate that gravitational waves propagate with the same speed of light in vacuo.
The extra numerical factor $2^2$, which can also be seen in $\frac{2\kappa c^3}{\kappa c^4/2}=\frac{4}{c}$ in parallel to the ratio of $\mu_0\epsilon_0$ in (\ref{MaxwellPE}), is due to the fact that gravity is a spin-2 instead of spin-1 field.
The minus sign $-\frac{\kappa c^4}{2}\rho_m$ in (\ref{GEMa}) compared with $\frac{\rho_e}{\epsilon_0}$ in (\ref{MaxwellPE}), the Gauss law,
reflects the fact that the ``charges" of the same sign in gravity attract rather than repel to each other.
The geodesic equation $\frac{du^\mu}{d\tau}+\Ga^\mu_{~\rho\si}u^\rho u^\si=0$ can be also put into the form \cite{Mashhoon:2003ax}
\bea\label{GraLorenf}
\frac{d\vec{v}}{dt}=\frac{m}{m}\left[\vec{E}_g(1+\frac{\vec{v}^2}{c^2})+\frac{\vec{v}}{c}\times\vec{B}_g\right]
\eea
analogous to the Lorentz force law, Eq. (\ref{LorentzF}).
In this analogy, gravitational mass can be regarded as the charge responsible for gravito-electromagnetic (GEM)
field, and weak equivalence principle ensures that the ``charge-to-mass" ratio is unit.
In fact, substituting $h_{0j}=\epsilon_{ijk}x^j\omega^k/c$ for an observer sticked to a rotating non-inertial frame in Minkowski spacetime into (\ref{GraLorenf}), the corresponding force $2m\,\vec{\omega}\times\vec{v}$ is exactly the Coriolis force, confirming that the non-inertial force and gravity may have the common origin, which is partially encoded in the Mach principle.
However, we have to keep caution that the formal analogy cannot be extended too far, though it proves quite fruitful, such as the prediction of gravito-magnetic precession of a spinning gyroscope in analogy with the magnetic dipole precession in magnetic fields, confirmed in Gravity Probe B project \cite{GPB1}\cite{GPB2},
and also in deriving solutions of the LV modified Einstein equation from the known ones in LV electrodynamics \cite{Bailey:2010af}.
The reason is that gravity is quite different from EM field:
{\it 1. the Maxwell equation is linear and EM field is abelian, while the Einstein equation is notoriously difficult to solve for its non-linearity;
2. the EM acceleration can be quite different for different particles with different charge-to-mass ratio,
while gravity is universal for all kinds of matter (attractive except for the cosmological constant \cite{PadmanabhanCC}) due to the equivalence principle}. Thus gravity can be geometrized while EM force cannot.
Technically,
1. Maxwell equation and Lorentz force law are gauge invariant and thus we can chose any gauge we like. This is not true in the case of gravito-electromagnetism, where only a restricted class of gauge transformations $h_{\mn}\rightarrow h'_{\mn}=h_{\mn}+2\prt_{(\mu}\xi_{\nu)}$ with $\prt^2\xi^\nu=0$ (satisfying the harmonic gauge) are allowed, otherwise the Maxwell-like equations (\ref{GEMa}) cannot hold.
2. The equations (\ref{GEMa}) and (\ref{GraLorenf}) are essentially not gauge invariant due to the two-layer structure of gravity: {\it the metric $g_{\mn}$ can be viewed as the potential of the connection $\Ga^\al_{\be\ga}$}, just as the definitions of $\vec{E}_g,~\vec{B}_g$ express (in this sense, equations (\ref{GEMa}), (\ref{GraLorenf}) are gauge invariant);
{\it while the connection $\Ga^\al_{\be\ga}$ is again the potential of the Riemann tensor $R^\la_{~\rho\mn}$, and the latter is the intrinsically ``gauge invariant" field strength.}
In other words, by working in the observer's local inertial frame or the Riemann normal coordinates,
we can always gauge away the force $m\frac{d\vec{v}}{dt}$ (derived from $m\frac{du^\mu}{d\tau}$).
In this respect, a set of essentially gauge invariant Maxwell-like equations must be based on
equations with covariant tensor forms, such as the Einstein equation and the geodesic deviation equation \cite{TidalEManalogy}.
The bonus of this choice is that we can go beyond linear approximations, and
the corresponding equations are more robust for further applications.
A detailed discussion of the essentially gauge invariant gravitational analogy of Maxwell electrodynamics
in the context of LV will be very interesting, however, this is beyond the scope of our present investigation.

\section{The fermion energy-momentum tensor in flat spacetime}\label{LVEMT}
The EM analogy in weak gravity is very useful cause electrodynamics is more easy and intuitive to deal with,
and we are more familiar with it, so we expect the similarity also arises between fermion-gravity (FG) and fermion-electromagnetic (FE) couplings.
The usual minimal FE coupling is in the form of $A_\mu {j_e}^\mu$,
where ${j_e}^\mu=-e\bar{\psi}\ga^\mu\psi$ is the conserved current.
The conservation is ensured by the gauge invariance of the FE coupling under gauge transformation $A_\mu\rightarrow A_\mu+\prt_\mu\Lambda$.
Similarly, in the weak field limit, we expect the minimal FG coupling takes a similar form
$-\hf h_{\mn}\Theta^{\mn}$,
where $\Theta^{\mn}$ is the symmetric energy momentum tensor (EMT).
In fact, from the gravitational definition of EMT \cite{WeinbergGrav},
\bea\label{GravEMT}
\Theta^{\mn}(x)\equiv\frac{2\delta I_\mathrm{M}}{\sqrt{-g(x)}\delta g_{\mn}(x)},
\eea
for a gauge transformation $\delta g_{\mn}(x)=2\nabla_{(\mu}\xi_{\nu)}$, the matter action $I_\mathrm{M}$ in (\ref{GravEMT}) transforms as
\bea&&
\delta I_\mathrm{M}=\hf\int d^4x\sqrt{-g}\,\delta g_{\mn}\Theta^{\mn}
\nn&&
=-\int d^4x\,\xi_\nu\sqrt{-g}\left\{\frac{\prt_\mu[\sqrt{-g}\Theta^{\mn}]}{\sqrt{-g}}+\Ga^\nu_{\mu\rho}\Theta^{\mu\rho}\right\}\nn&&~~~~+\int d^4x\,\prt_\mu\left[\sqrt{-g}\xi_{\nu}\Theta^{\mn}\right],
\eea
where the terms in the large brace above is exactly $\nabla_{\mu}\Theta^{\mn}$.
Ignoring the surface term $\sqrt{-g}\xi_{\nu}\Theta^{\mn}$, gauge invariance again ensures the covariant conservation of EMT, $\nabla_{\mu}\Theta^{\mn}=0$.
Unlike the case of EM matter couplings, there is no simple conservation law of EMT $\prt_\mu\Theta^{\mn}=0$
for the case of gravity, though the linear gauge transformation $\delta h_{\mn}=2\prt_{(\mu}\xi_{\nu)}$
may lead to the ordinary current conservation for the coupling $-\hf h_{\mn}\Theta^{\mn}$.
This is quite similar to the non-abelian Yang-Mills theory,
where no simple conservation law exist for a current constructed purely from matter field,
$J_a^{~\nu}=-i\frac{\prt\mathcal{L}_\mathrm{M}}{\prt D_\nu\psi}t_a\psi$ \cite{WeinbergQFT2}.
$J_a^{~\nu}$ is only covariantly conserved, $D_\nu J_a^{~\nu}=0$.
To construct an ordinary conserved EMT $\prt_\mu\tau^{\mn}=0$, just like the ordinary conserved current $\mathscr{J}_a^\mu\equiv J_a^{~\mu}-C^c_{~ab}F_c^{\mn}A^b_\nu$ ($\prt_\mu\mathscr{J}_a^\mu=0$) contains contribution from non-abelian gauge field itself,
the ordinary conserved EMT $\tau^{\mn}$ must also contain contribution from gravitational field itself,
\ie, terms proportional to the summation of powers of metric tensors and their derivatives,
such as the Landau-Lifschitz pseudotensor $t_\mathrm{LL}^{\mn}$, then $\tau^{\mn}=(-g)\left[T^{\mn}+t_\mathrm{LL}^{\mn}\right]$ \cite{CTFLL}.
In other words, the gravitational field itself carries energy and momentum, and thus contributes to the source of gravity.
This has already been dramatically verified by the direct observation of gravitational waves \cite{GW2015}.
In fact, the stress-energy tensor for the GEM field in stationary case can be shown to be exactly proportional to the pseudotensor $t_\mathrm{LL}^{\mn}$ \cite{Mashhoon:2003ax}.

In flat spacetime, the canonical formalism gives another way to obtain EMT as the zero-gravity limit of the matter ``source current" for gravity,
provided the Belinfante symmetrization procedure (B-procedure) is performed.
However, in the presence of LV, the usual Belinfante symmetrization may not be attainable \cite{SMEc}.
As an example, consider the following SME Lagrangian \cite{SMEc},
\bea\label{FerGauLV}&&
\mathscr{L}=\mathscr{L}_0+\delta\mathscr{L}_\mathrm{LV},\nn&&
\mathscr{L}_0=\frac{i}{2}\psb\gamma^\mu\lrdrt\mu\ps-m_\psi\psb\psi-\hf\mathrm{Tr}[F^{\mn}F_\mn],\nn&&
\de\cL_\mathrm{LV}=\frac{i}{2}\psb\de\Gamma^\mu\lrdrt\mu\ps-\psb\de{M}\ps,\nn&&
\bar{\chi}\Ga^\mu\lrdrt\mu\ps\equiv\,\bar{\chi}\Ga^\mu{D}_\mu\ps-\bar{\chi}\bar{D}_\mu\Ga^\mu\ps,
\eea
where $\bar{\chi}\bar{D}_\mu\Ga^\mu\ps\equiv[(\prt_\mu-ieA_\mu)\bar{\chi}]\Ga^\mu\ps$,
$\de\Gamma^\mu\equiv\Gamma^\mu-\ga^\mu\equiv-\left[c^{\nu\mu}\ga_\nu+d^{\nu\mu}\ga_5\ga_\nu\right]$ and
$\de{M}\equiv a^\mu\ga_\mu+b^\mu\ga_5\ga_\mu+\hf{H}^\mn\si_\mn$.
Note for simplicity, the $e^\mu,~f^\mu,~g^{\la\mn}$ coefficients are dropped.
Except the $c^{\nu\mu}$ and $a^\mu$, all the other LV coefficients are responsible for the LV spin-interactions \cite{arbitrFermion}.
We include $a^\mu$ term as in the presence of gravity, the $a^\mu$ coefficient cannot be totally removed
by field redefinition even for fermions with a single flavor, unlike the case of flat spacetime \cite{Prospects2008}.
Also note there is a sign difference for the $c,~d$ coefficients in $\Ga^\mu$,
as the signature for Minkowski metric is $\eta_{\mn}=\mathrm{diag}(-1,+1,+1,+1)$,
the one conventionally adopted in the gravity community, rather than the one in QFT \cite{SMEa}.
Only in this section, we use Greek indices to denote variables in Minkowski spacetime,
while in the next following sections,
we use Latin indices $a,~b,~c...$ from beginning for tagent space variables
and the Latin indices $i~,~j,~k,...$ in the middle for pure spatial indices,
while the Greek indices $\mu,~\nu,~\rho,...$ are for manifold variables.
Similarly, the convention for the totally antisymmetric tensor is fixed by $\epsilon_{0123}=1$.

From the Lagrangian (\ref{FerGauLV}), we get the canonical momenta from the definition $\Pi_l\equiv{\prt\mathscr{L}}/{\prt\dot{\Psi}^l}$,
\bea\label{CanoMLV}&&
\Pi^\mu_\psi\equiv\frac{i}{2}\psb[\gamma^\mu+\de\Gamma^\mu],  \quad \Pi^\mu_{\psb}\equiv-\frac{i}{2}[\gamma^\mu+\de\Gamma^\mu]\psi,\nn&&
-\Pi^\mu_{A^a_{~\rho}}\equiv\,F^{a\mu\rho}=\prt^\mu{{A^a}^\rho}-\prt^\rho{{A^a}^\mu}+f^a_{~bc}{A^b}^\mu{A^a}^\rho.
\eea
The canonical EMT denoted as $T^{\mn}$ is obtained as below
\bea\label{CanoEMTLV}&&
\hspace{-3mm}T^{\mn}\equiv\Pi^\mu_\psi\,\partial^\nu\psi+\partial^\nu\psb\,\Pi^\mu_{\psb}+\Pi^\mu_{A^a_{~\rho}}\partial^\nu{A}^a_{~\rho}
-\eta^{\mn}\mathcal{L}\nn&&~~~~={T_0}^{\mn}+\de{T}^{\mn},\nn&&
\hspace{-3mm}{T_0}^{\mn}\equiv\frac{i}{2}\left[\psb\gamma^\mu\prt^\nu\psi-(\partial^\nu\psb)\gamma^\mu\psi\right]
-F^{a\mu\rho}\partial^\nu{A}^a_{~\rho}\nn&&~~~~~~
-\eta^{\mn}\mathcal{L}_0,\nn&&
\hspace{-3mm}\de{T}^{\mn}\equiv\frac{i}{2}\left[\psb\de\Gamma^\mu\partial^\nu\psi-(\partial^\nu\psb)\de\Gamma^\mu\psi\right]
-\eta^{\mn}\de\mathcal{L}_\mathrm{LV}.
\eea
In the absence of gravity, the violation of Lorentz invariance does not conflict with the spacetime translation invariance,
which is assumed to be hold since we do not want to loss the energy-momentum conservation, $\prt_\mu T^{\mn}=0$,
provided the fields and their derivatives vanishes sufficiently quickly at spatial infinity.
However, the B-procedure does not work, as it crucially relies on the fact that the total angular momentum tensor density is conserved, $\prt_\mu\mathscr{J}^\mu_{~\al\be}=0$, where
\bea\label{CGLC}
\mathscr{J}^\mu_{~\al\be}\equiv\frac{\prt\cL}{\prt[\prt_\mu\Ps(x)]}\left[S_{\al\be}\right]\Ps(x)+x_\al{T}^\mu_{~\be}-x_\be{T}^\mu_{~\al}\nonumber
\eea
includes the intrinsic spin contribution $\mathcal{S}^\mu_{~\al\be}\equiv\frac{\prt\cL}{\prt[\prt_\mu\Psi(x)]}\left[S_{\al\be}\right]\Psi(x)$ due to
the non-trivial field representation of the Poincar$\acute{\mathrm{e}}$ group.
Since Lorentz invariance is broken, the total angular momentum needs not to be conserved, $\prt_\mu\mathscr{J}^\mu_{~\al\be}\neq0$.
To see why LV blocks the construction of a symmetric EMT, first we note that the antisymmetric part of the canonical EMT is
\bea
T^{[\al\be]}\equiv\hf\left(T^{\al\be}-T^{\be\al}\right)=\hf\prt_\mu\left[\mathscr{J}^{\mu\al\be}-\mathcal{S}^{\mu\al\be}\right],
\eea
where we have used $\prt_\al T^{\al\be}=0$.
Now suppose adding $T^{\al\be}$ with a total derivative $\prt_\rho\mathscr{A}^{\rho\al\be}$,
provided that $\mathscr{A}^{\rho\al\be}$ vanishes sufficiently fast at spatial infinity.
Then the improved EMT is
\bea&&
\hspace{-8mm}\Theta^{\al\be}=T^{\al\be}+\prt_\rho\mathscr{A}^{\rho\al\be}\nn&&
=T^{(\al\be)}+\prt_\rho\left\{\hf\left[\mathscr{J}^{\rho\al\be}-\mathcal{S}^{\rho\al\be}\right]+\mathscr{A}^{\rho\al\be}\right\},
\eea
where the unaltered conservations law requires $\mathscr{A}^{\rho\al\be}=-\mathscr{A}^{\al\rho\be}$.
Clearly, still adopting the Belinfante-Rosenfield formalism \cite{Belin1939}\cite{Rosen1940}
and letting
\bea\label{BelinA}
\mathscr{A}^{\rho\al\be}\equiv\mathcal{S}^{\rho\al\be}-\mathcal{S}^{\al\rho\be}-\mathcal{S}^{\be\rho\al},\nonumber
\eea
we can confirm that $\mathscr{A}^{\rho\al\be}=-\mathscr{A}^{\al\rho\be}$ from its definition.
The B-procedure indeed guarantees the current conservation $\prt_\al\Theta^{\al\be}=0$, only
in general $\Theta^{\al\be}\neq\Theta^{\be\al}$ due to the presence of Lorentz violation.
This feature of EMT has been clarified with an explicit example in \cite{SMEa}, and has
been discussed in depth including gravity in the Rieman-Cartan geometry \cite{SMEc}.

As Lorentz violation forbids a conserved angular momentum current, we do not expect a natural symmetric EMT.
Moreover, the B-procedure cannot even necessarily give rise to a gauge invariant EMT.
This has been observed in the LV modified electromagnetism with the $k_{AF}$ term \cite{SMEa} already.
As another example, we show after B-procedure, the EMT with only $c$-coefficient is
\bea\label{EMTc-coe}&&
\hspace{-6mm}\Theta^{\mn}\equiv\Theta^{\mn}_\ps+\Theta^{\mn}_A\nn&&
=\frac{i}{4}\psb\left[(\ga^\nu\lrdut\mu+\ga^\mu\lrdut\nu)\right]\ps-F^{\mu\rho}F^\nu_{~\rho}-\eta^{\mn}\mathcal{L}_0\nn&&
+\frac{i}{4}\psb\left[c_\rho^{~\nu}(\ga^\rho\lrdut\mu-\ga^\mu\lrdut\rho)
-c_\rho^{~\mu}(\ga^\rho\lrdut\nu+\ga^\nu\lrdut\rho)\right.\nn&&
\left.+(\ga^\mu c^\nu_{~\rho}-\ga^\nu c^\mu_{~\rho})\lrdut\rho\right]\ps
+e\psb\left[(c_\rho^{~\mu}A^\nu+c_\rho^{~\nu}A^\mu)\ga^\rho\right.\nn&&
\left.
-(\ga^\nu c_\rho^{~\mu}+\ga^\mu c_\rho^{~\nu})A^\rho\right]\ps-\eta^{\mn}\delta\mathcal{L}_\mathrm{LV},
\eea
where we have ignored 2nd order LV corrections. Clearly, the terms proportional to $c$-coefficients block the
symmetrization, $\Theta^{[\mn]}\neq0$, and the terms in the third square bracket even block the gauge invariance.
Without LV, the terms in the second line is manifestly symmetric and gauge invariant,
and coincide with the gauge invariant EMT, Eq. (4-5) for quark and gluon in \cite{GaugeIXdJ1},
up to a signature difference.
Interestingly, the B-procedure does give a gauge invariant EMT for pure LV gauge field with Lagrangian
\bea\label{LIGaugeL}
\mathscr{L}_{A}=-\frac{1}{4}\left[{F^a}^{\mn}{F^a}_{\mn}+(k_F)^{\mn\rho\si}{F^a}_{\mn}{F^a}_{\rho\si}\right],
\eea
where ${F^a}_{\mn}\equiv\prt_\mu {A^a}_\nu-\prt_\nu {A^a}_\mu+f^a_{~bc}{A^b}_\mu{A^c}_\nu$ is the field strength
for the gauge field $A^a$. The B-procedure improved EMT is
{\small
\bea
\hspace{-6mm}\Theta_A^{\mn}=-{F^a}^\mu_{~\kappa}{F^a}^{\nu\kappa}
-(k_F)^{\mu~\al\be}_{~\kappa}{F^a}_{\al\be}{F^a}^{\nu\kappa}-\eta^{\mn}\mathscr{L}_{A},
\eea
}
\hspace{-2mm}which is apparently gauge invariant, but still not symmetric.
In view of these examples, we see that to have a gauge invariant improvement of the canonical EMT,
seems other improvement procedures are required rather than Belinfante symmetrization procedure,
which is even not attainable.
Not only because symmetrization is blocked by the presence of LV, which is equivalent to the presence
of background tensor fields causing the asymmetry, but also because
symmetrization is only indicated by the metric framework of gravitational theory.
For a generic gravitational theory allowing other degrees of freedom (d.o.f.), such as torsion or non-metricity \cite{GRwithST}\cite{SMEc},
the generalized Einstein equation does not require a symmetric EMT as the source of gravity,
though an effective symmetric EMT is always attainable if we separate the Einstein tensor into the Riemannian part,
and incorporate the non-Riemannian part into the effective EMT \cite{GRwithST}.
However, the cost is that it plagues a proper interpretation of the gravitation and matter d.o.f.
We will postpone a further investigation of LV EMT in the future,
and turn to discuss FG couplings in the next section.

\section{Preliminary for fermion-gravity interactions}\label{NRMEtriC}
To consider the fermion-gravity couplings, the flat space LV fermion Lagrangian (\ref{FerGauLV})
has to be replaced by the curved space version \cite{SMEc}\cite{TassonMGC}
\bea\label{matterAct}&&
\hspace{-5mm}\mathscr{L}_\psi=e\left[\frac{i}{2}e^\mu_{~a}\bar{\psi}\Ga^a\,{\lrNrt\mu}\,\psi-\bar{\psi}M\psi\right],\\\label{CovariD}&&
\hspace{-5mm}\psb\Ga^a\,{\lrNrt\mu}\psi\equiv\psb\Ga^a[\vec{D}_\mu+\frac{i}{4}\omega_\mu^{~bc}\si_{bc}]\ps
-\psb[\overleftarrow{D}_\mu-\frac{i}{4}\omega_\mu^{~bc}\si_{bc}]\Ga^a\ps,\nn\label{GGamma}&&
\hspace{-5mm}\Ga^a\equiv\ga^a-\left[c_{\rho\nu}\ga^b+d_{\rho\nu}\ga_5\ga^b\right]e^{\nu a}e^\rho_{~b},\\\label{GMass}&&
\hspace{-5mm}M\equiv{m+a_\mu e^\mu_{~a}\ga^a+b_\mu e^\mu_{~a}\ga_5\ga^a+\hf H_{\mn}e^\mu_{~a}e^\nu_{~b}\si^{ab}},
\eea
Note we use $\vec{D}_\mu\ps=(\prt_\mu+igA_\mu)\ps$ to represent pure gauge coupling
and $\vec{\nabla}_\mu\psi=[\vec{D}_\mu+\frac{i}{4}\omega_\mu^{~bc}\si_{bc}]\ps$ to represent the covariant derivatives including both the minimal gauge field (the gauge field means photon in this context) and the spin connection couplings.
Also note that we use $g$ instead of $e$ to represent gauge coupling to avoid confusion with the determinant of vierbein, as it will be more easier to distinguish determinant of metric from the coupling constant $g$ in this context.
The gravity sector is assumed to be intact to largely simplify the original construction with torsion
in Riemann-Carton spacetime \cite{SMEc}.
We mention that torsion and non-metricity can also be tightly constrained in the context of SME \cite{TorsSME}\cite{TorGalaxy}\cite{NonMetSME}, though they draw great attention to the gravity community
even in the LI context \cite{GRwithST}\cite{TorShapiro}\cite{DiraMajXX}.
Considering the weak gravity limit up to the lowest order of metric perturbation $h_{\mn}=g_{\mn}-\eta_{\mn}$,
the vierbein and spin connection can thus be written as
\bea\label{tetradSC}&&
\hspace{-3.5mm}e_\mu^{~a}\simeq\delta_\mu^{~a}+\hf h_\mu^{~a}+\chi_\mu^{~a},\quad
e_{~a}^{\mu}=\delta_{a}^{\mu}-\hf h_{~a}^{\mu}+\chi_{~a}^{\mu},\nn&&
\hspace{-3.5mm}\omega_\mu^{~a b}=\hf\left[e^{\nu a}(\prt_\mu e_\nu^{~b}-\prt_\nu e_\mu^{~b})-e_\mu^{~c}\prt_\al e_{\be c}e^{\al a}e^{\be b}\right]-(a\leftrightarrow b)\nn&&~~
\simeq \hf(h_{\mu\,,}^{~a\,b}-h_{\mu\,,}^{~b\,a})+\chi^{ab}_{~~,\mu}+\chi_{\mu\,,}^{~a\,b}-\chi_{\mu\,,}^{~b\,a},
\eea
The $\chi_{ab}=-\chi_{ba}$ contain the 6 local Lorentz degrees of freedom in the vierbein,
and can be totally removed by fermion field redefinition $\psi(x)\rightarrow\exp[-\frac{i}{4}\chi_{ab}(x)\si^{ab}]\psi(x)$ \cite{TassonMGC}\cite{Bluhm:2004ep}.
This redefinition may still leave imprints on the fluctuations of LV coefficients\cite{TassonMGC},
however, due to stringent experimental constraints \cite{dataTable} and our solely interest in the effects
caused by the vacuum expectation values (vev) of the LV coefficients,
we can safely ignore $\chi$ in the following. For example,
\bea&&
e^\mu_{~a}\Ga^a=\ga^\mu-\hf h_\mu^{~a}\ga^a-[c_b^{~\mu}-c_{b\nu}h^{\mn}-\hf c_\rho^{~\mu}h^\rho_{~b}]\ga^b\nn&&~~~~
-[d_b^{~\mu}-d_{b\nu}h^{\mn}-\hf d_\rho^{~\mu}h^\rho_{~b}]\ga_5\ga^b.
\eea
Note different from the notation in \cite{TassonMGC}, we use mixed Latin and Greek indices to keep track of their origin, though all the indices can be put into the Greek ones, since we take $h_{\mn}$ as the metric deviation from the vacuum Minkowski background.
In other words, in the following discussions of linearized weak gravity,
there is no need to distinguish Latin and Greek indices, as all the upper and lower indices,
whether Latin or Greek, are raised or lowered by the corresponding Minkowski metric.
As the LV coefficients are linear on the level of Lagrangian, we can treat them one by one.
First note we can separate the Lagrangian (\ref{matterAct}) into LI and LV parts,
$\mathscr{L}_\psi=(1+\hf h)\left[\mathscr{L}_\mathrm{LI}+\mathscr{L}_\mathrm{LV}\right]$,
where the determinant of the vierbein $e=\sqrt{-g}=1+\hf h$.
The LI Lagrangian can be written as
\bea\label{QEDLIGrav}&&
\mathscr{L}_\mathrm{LI}=\frac{i}{2}e^\mu_{~a}\bar{\psi}\ga^a\,{\lrNrt\mu}\psi-\bar{\psi}m\psi\nn&&
\simeq\frac{i}{2}\psb\left[\ga^a\lrdrt a-\hf h^\mu_{~a}\ga^a\lrdrt\mu\right]\psi-\bar{\psi}m\psi,
\eea
where the ``$\simeq$" means preserving only terms up to linear order of $h_{\mn}$, and we have utilized the identity $h_{ab,c}\{\ga^a,~\si^{bc}\}=0$.
The LV counterpart is
\bea\label{QEDLVGrav}&&
\mathscr{L}_\mathrm{LV}=\frac{i}{2}e^\mu_{~a}\bar{\psi}\delta\Ga^a\,{\lrNrt\mu}\,\psi-\bar{\psi}\delta M\ps\simeq
\nn&&
\hspace{-5mm}\frac{i}{2}\bar{\psi}\left[\delta\Ga_\circ^a\left(\delta^\mu_{~a}-\hf h^\mu_{~a}\right)\lrdrt\mu+\delta\Ga_h^a\lrdrt a\right]\ps-\bar{\psi}\left(\delta M_\circ+\delta M_h\right)\psi\nn&&
\hspace{-5mm}+\frac{1}{4}\epsilon^{bcmn}h_{am,n}\psb[c_b^{~a}\ga_5\ga_c+d_b^{~a}\ga_c]\ps,
\eea
where for simplicity, we have defined
{\small
\bea&&
\delta\Ga_h^a\equiv\hf\left[h^{\nu a}(c_{b\nu}+d_{b\nu}\ga_5)+h^\rho_{~b}(c_{\rho}^{~a}+d_{\rho}^{~a}\ga_5)\right]\ga^b, \nn&& \delta\Ga_\circ^a\equiv-\left[c_b^{~a}\ga^b+d_b^{~a}\ga_5\ga^b\right],\nn&&
\delta M_h\equiv-\hf h^\mu_{~a}[(a_\mu+b_\mu\ga_5)\ga^a+\hf H_{\mu b}\si^{ab}]-\frac{1}{4} h^\nu_{~b}H_{a\nu}\si^{ab},\nn&&
\delta M_\circ\equiv(a_a\ga^a+b_a\ga_5\ga^a+\hf H_{ab}\si^{ab}).\nonumber
\eea
}
\hspace{-1.5mm}To the linear order of metric perturbation, the Euler-Lagrangian equation with respect to
$\mathscr{L}_\mathrm{LI}+\mathscr{L}_\mathrm{LV}$ is
{\small
\bea\label{LVDiraclinear}&&
\hspace{-5mm}\left\{i\left[(\Ga^a_\circ+\delta\Ga^a_h)\vec{D}_a-\frac{h^\mu_{~a}}{2}\Ga^a_\circ\vec{D}_\mu\right]
-(M_\circ+\delta{M}_h)\right\}\psi+\frac{i}{2}{\Big[}\prt_a\delta\Ga^a_h\nn&&
\left.-\hf\prt_ah^a_{~c}\Ga^c_\circ
-\frac{i}{2}\epsilon^{bcd}_{~~~e}h_{ab,c}(c_d^{~a}\ga_5+d_d^{~a})\ga^e\right]\psi=0.
\eea
}
\hspace{-1.5mm}where $\Ga_\circ^a=\ga^a+\delta\Ga_\circ^a$ and $M_\circ=m+\delta M_\circ$.
Note we haven't considered the so-called geometric term
\bea\label{geoC}
\mathscr{L}_\mathrm{geo}=(e-1)\mathscr{L}_\psi\simeq
\frac{h}{2}\left[\frac{i}{2}\bar{\psi}\Ga_\circ^a\,{\lrdrt a}\,\psi-\bar{\psi}M_\circ\psi\right],
\eea
since this term comes from the artifact of linearization, which amounts to nothing but multiplying Eq. (\ref{LVDiraclinear}) with a rescaling factor $e=1+\frac{h}{2}$.
If pick up back these terms, the equation is exactly the one obtained from the linearization of the full Dirac equation with respect to the Lagrangian (\ref{matterAct}).

In comparison with the EM coupling $-j^\mu A_\mu$, we can also collect all the terms proportional to $h_{\mn}$
in the Lagrangian, which is
\bea&&\label{hIntGM}
\hspace{-3mm}\mathscr{L}_\mathrm{hI}=\nn&&
-\hf h^\mu_{~a}\left\{\frac{i}{2}\psb\Ga_\circ^a\lrdrt\mu\ps-\psb\left[(a_\mu+b_\mu\ga_5)\ga^a
+H_{\mu b}\si^{ab}\right]\ps\right\}\nn&&
+\frac{i}{4}\bar{\psi}\left[h^{\nu a}(c_{b\nu}+d_{b\nu}\ga_5)
+h^{\rho}_{~b}(c_\rho^{~a}+d_\rho^{~a}\ga_5)\right]\ga^b\lrdrt a\ps\nn&&
+\frac{1}{4}\epsilon^{mnb}_{~~~~c}h_{am,n}\psb[c_b^{~a}\ga_5+d_b^{~a}]\ga^c\ps\nn&&
+\frac{h}{2}\left[\frac{i}{2}\bar{\psi}\Ga_\circ^a\,{\lrdrt a}\,\psi-\bar{\psi}M_\circ\psi\right]
\dot{=}-\hf h_{\mn}T^{\mn},
\eea
where all the Greek and Lain indices are raised or lowered by the corresponding Minkowski metric,
and thus loss the distinctive features they have before the linearization.
Also note ``$\dot{=}$" means equal up to a total derivatives, since we have dropped a total derivative term proportional to $\epsilon^{\mu bcd}$,
and the energy-momentum tensor $T^{\mn}$ is given explicitly,
\bea\label{EMTlin}&&
T^{\mn}=\frac{i}{2}\psb\Ga_\circ^\nu\lrdut\mu\ps-\psb\left[(a^\mu+b^\mu\ga_5)\ga^\nu+H^{\mu}_{~b}\si^{\nu b}\right]\ps\nn&&
~~
-\frac{i}{2}\psb\left[(c_b^{~\mu}+d_b^{~\mu}\ga_5)\ga^b\lrdut\nu+(c^\mu_{~b}+d^\mu_{~b}\ga_5)\ga^\nu\lrdut b\right]\ps\nn&&
~~-\eta^{\mn}\mathscr{L}_\mathrm{flat}+i\epsilon^{\mu bcd}\prt_d\left[\psb(c_b^{~\nu}\ga_5+d_b^{~\nu})\ga_c\ps\right],
\eea
where $\mathscr{L}_\mathrm{flat}\equiv\frac{i}{2}\bar{\psi}\Ga_\circ^a\,{\lrdrt a}\,\psi-\bar{\psi}M_\circ\psi$.
Aside from the last term coming from spin-connection, the EMT obtained in this way is gauge invariant and symmetric,
as the apparently asymmetric part $T^{[\mn]}$ does not contribute due to the coupling with $h_{\mn}=h_{\nu\mu}$.
The LV coefficients in the above are just vevs and thus are spacetime independent.

Now the form $\mathscr{L}_\mathrm{hI}\dot{=}-\hf h_{\mn}T^{\mn}$ is similar to the EM coupling $-A_\mu j^\mu$,
and can be regarded as a linear approximation of $\delta I=\hf\sqrt{-g}\delta g_{\mn}T^{\mn}$ up to the determinant $\sqrt{-g}$.
It is interesting to note that the geometric contribution in (\ref{geoC}) cannot be ignored as stated in \cite{Adler2011bg}, otherwise
the resultant EMT will differ by a term proportional to $\eta^{\mn}$ compared to the EMT obtained with the canonical formalism.
However, this term doesn't contribute if the matter fields are on the mass shell since we only consider metric couplings up to linear order.

As mentioned already, we can study the non-relativistic fermion-gravity interaction from the well-known FW transformation method \cite{FWT1950}\cite{SemiClaGF-Silenko}\cite{Obukhov2009},
which requires a relativistic Hamiltonian with conventional time evolution as the starting point.
As our main concern, the other way is to calculate the interaction energy between a pair of one-fermion states,
$\int d^3\vec{x}\langle p',\be|-\mathcal{L}_\mathrm{int}|p,\al\rangle=\hf\int d^3\vec{x}\,h_{\mn}\langle p',\be|T^{\mn}|p,\al\rangle$,
where $\langle p',\be|T^{\mn}|p,\al\rangle$ is the gravitational form factor
extensively studied in hadron spin structures \cite{FGEMT-Jixd1995}.
However, even for the latter approach, to find out the proper eigen-spinors
for proper Fourier expansion of $\psi(x)$, we still face the same necessity of field redefinition.
In fact, even for a covariant Dirac equation without unconventional time derivatives impeding
the proper identification of the time evolution operator \cite{TassonMGC},
field redefinition is still an essential step to get a hermitian Hamiltonian \cite{Obukhov2009},
and has been well developed in the context of SME \cite{SMEa}\cite{TassonMGC}\cite{arbitrFermion}
to study perturbative LV effects, such as the effects due to LV fermion-gravity couplings.
We will discuss the field redefinition in a proper place in the next section.

\section{Non-relativistic fermion-gravity coupling and the analogy}\label{Analogy}
The method to get NR interaction energies is adopted from Ref. \cite{Adler2011bg},
where the basic idea is from the lessons we learn in QED.
In QED, the electrostatic force is mediated by the photon exchange between two charged particles, and the full relativistic interaction is described by the vector-current interaction $-j^\mu A_\mu$ with $j^\mu=\psb\Ga^\mu\ps$,
where $\Ga^\mu=\ga^\mu$ in LI QED.
Likewise, the gravitational interaction is mediated by the graviton exchange between two energy carriers
(without NR approximation, massless particles such as photon are also allowed),
and the full relativistic interaction is described by the tensor-current interaction $-\hf h_{\mn}T^{\mn}$,
where $T^{\mn}$ is given by Eq. (\ref{GravEMT}) in general, and only the symmetrized part of $T^{\mn}$ really contributes.
For the Lagrangian (\ref{matterAct}), $T^{\mn}$ is explicitly given by (\ref{EMTlin}) in the linearized approximation.
Follow the same logic, we try to get the leading order NR one-fermion interaction matrix elements
from the fully relativistic interaction Lagrangian (\ref{hIntGM}).

In standard QFT, the spinor $\ps$ can be expanded as
{\small\bea&&\label{fermion}
\hspace{-8.5mm}\ps(x)=\sum_{\si=1,~2}\int\tilde{dk}\left[\hat{b}_\si(\vec{k})u_\si(\vec{k})e^{ik\cdot x}
+\hat{d}_\si^\dagger(\vec{k})v_\si(\vec{k})e^{-ik\cdot x}\right],
\eea}
\hspace{-1.5mm}where $d\tilde{k}\equiv\frac{d^3k}{(2\pi)^3}\frac{m}{k^0}$, $k\cdot x\equiv\vec{k}\cdot\vec{x}-k^0x^0$,
and $u_\si,~v_\si$ are the eigen-spinors describing electron and positron, respectively.
In the LI situation, the explicit forms of $u_\si,~v_\si$ can be found in any text books of QFT, say \cite{QFTinNS}\cite{QFT-Itzykson}.
However, in the presence of generic LV couplings, the physical free-particle states cannot be directly described by $\psi$ due to unconventional time evolution imposed by the LV derivative couplings, such as the $c,~d$ terms.
To eliminate the extra time-derivatives, we have to invoke the spinor redefinition $\psi=\hat{U}\chi$
to cast the kinematic term into the conventional structure $\hf i\bar{\chi}\ga^0\lrprt0\chi$,
which also preserves the usual scalar product $\langle\Psi|\Phi\rangle=\int d^3x\,\Psi^\dagger\Phi$ \cite{TassonMGC} in flat space.

For the flat space Lagrangian (\ref{FerGauLV}), $\hat{U}=(\ga^0\Ga_\circ^0)^{-\hf}$ is a non-singular spacetime-independent matrix \cite{Lehnert2004}.
For a generic fermion Lagrangian, the redefinition matrix is given by Eq.(30) in Ref. \cite{TassonMGC}
up to leading order of perturbative parameters of $h_{\mn}$ and LV coefficients.
Thus in general $\hat{U}$ can be quite complicated and spacetime-dependent.
The explicit form of $\hat{U}$ corresponding to Lagrangian (\ref{matterAct}) is given in Appendix \ref{FieldRedP},
and can be shown to satisfy $\hat{U}^\dagger\ga^0\Ga^0\hat{U}=\hat{I}$ \cite{Lehnert2001}.
As what we concerned is $\mathscr{L}_\mathrm{hI}$ in Eq.(\ref{hIntGM}), it suffices to use
the flat space redefinition matrix $\hat{U}_0\equiv1+\hf(d_{b0}\ga_5-c_{b0})\ga^0\ga^b$ for a linear approximation.
In fact, detailed calculations lead to additional $h$-couplings from the flat space Lagrangian
$\mathscr{L}^\mathrm{flat}_\psi=\frac{i}{2}\bar{\psi}\Ga_\circ^a\,{\lrdrt a}\,\psi-\bar{\psi}M_0\psi$, as
the spinor redefinition matrix $\hat{U}\supset\de{\hat{U}}^h$,
however, these terms do not contribute to $\langle p',\be|\int d^3x\mathscr{L}^\mathrm{flat}_\psi|p,\al\rangle$, provided the external fermions are on mass-shell.

For non-derivative LV couplings such as $a,~b,~H$ coefficients, the LI eigen-spinor may serve as first order approximation in Eq. (\ref{fermion}).
While for the $c,~d$ coefficients with extra time-derivatives, the quantization expansion in terms of $\hat{b}_\si,~\hat{d}_\si$ has to be done with redefined spinor $\chi$ directly.
Of course, the eigen-spinor can always be written as $S_\al=S_{\al0}+S_{\al1}$ ($S_\al$ refers to either
$u_\al$ or $v_\al$), where $S_{\al0},~S_{\al1}$ denote the LI and LV contributions, respectively.
For the 1-fermion matrix elements $\langle p',\be|\hat{O}|p,\al\rangle$
at leading order approximation, the key ingredient in the Fourier expansion can be written as
\bea\label{MatrixE1PS}&&
\hspace{-7mm}\bar{S}_\be\hat{O}S_\al=\bar{S}_{\be0}(\hat{O}_0+\hat{O}_1)S_{\al0}
+(\bar{S}_{\be1}\hat{O}_0S_{\al0}+\bar{S}_{\be0}\hat{O}_0S_{\al1}),\nn
\eea
where $\bar{S}_\al\equiv{S_\al}^\dagger\ga^0$ is the Dirac adjoint of the eigen-spinor $S_\al$,
$\hat{O}$ denotes any operator we are interested, such as $e\vec{A}\cdot\vec{\Ga}$,
and $\hat{O}_0,~\hat{O}_1$ denote the LI and LV separations of $\hat{O}$.
Since in the NR limit, the contribution from the spinor $v_\si(\vec{k})$ with negative energy can be totally ignored,
and the scattered fermion is assumed to be always on the mass shell, Eq. (\ref{fermion}) becomes
(for $c, d$ coefficients, $\psi$ has to be replaced by $\chi=\hat{U}^{-1}\psi$)
\small\bea&&\label{Fer-wavefun}
\hspace{-5mm}\ps(x)=\int\,d\tilde{k}\sum_{\si=1,~2}\hat{b}_\si(\vec{k})u_\si(\vec{k})e^{ik\cdot x},
\eea
where $k^0=k^0[\vec{k}, m, X]$ is the LV modified dispersion relation,
and $X$ represents a set of generic LV coefficients with indices suppressed.
The LI eigen-spinor is
\bea\label{LIES}
u_{\si0}(\vec{k})=\sqrt{\frac{\omega_0+m}{2m}}\left(
         \begin{array}{c}
           \xi^{\si} \\
           U_0(k)\xi^{\si} \\
         \end{array}
       \right)\stackrel{\mathrm{NR}}{\simeq}
       \left(
         \begin{array}{c}
           \xi^{\si} \\
           \frac{\vec{\sigma}\cdot\vec{k}}{2m}\xi^{\si} \\
         \end{array}
       \right),
\eea
where $U_0(k)\equiv\frac{\vec{\sigma}\cdot\vec{k}}{\omega_0+m}$ and $\omega_0=\sqrt{\vec{k}^2+m^2}$.
For $a,~b$ type coefficients, the eigen-spinor can be directly found in the appendix in \cite{SMEa}.
For completeness, we collect them together with the eigen-spinors for $c,~d,~H$ coefficients in Appendix \ref{LVeigenSpinors}.
The key idea is that since LV is supposed to be tiny by observational constraints \cite{dataTable},
we only need to {\it keep linear order LV corrections, and hence can treat various LV coefficients one by one
as if the other LV coefficients are absent}.
Thus in calculating LV contributions of FG or FE 
interaction energies from matrix elements, we can classify them into 3 categories:
\bit
\item Apparent LV vertices, such as $\mathcal{O}_\mathrm{LV}=\frac{h_{ba}}{2}\psb[(a^b+b^b\ga_5)\ga^a+H^b_{~c}\si^{ac}]\ps$,
where Eq. (\ref{fermion}) with LI eigen-spinors is sufficient;
\item LV eigen-spinor induced LV to the superficially LI vertices, such as $\mathcal{O}_\mathrm{LI}=\frac{ih_{ba}}{4}\psb\ga^a\lrdut b\ps$,
where $\ps$ and $\psb$ receive LV corrections and thus induce LV corrections to interaction energy.
In this case, the eigen-spinor correction appears through the LV corrected matrix connecting the upper and
lower two Pauli 2-spinor $\xi^\al$ and $U_X(k)\xi^\al$, \ie, $U_0(k)\rightarrow U_X(k)$, where $X$ again represents certain LV coefficient
with Lorentz indices suppressed;
\item LV correction to dispersion relations, which in the NR limit may also induce LV corrections,
such as $\frac{1}{E(p,X)+m}\simeq\frac{1}{2m}[1-\frac{X}{4m}]$, where $X$ represents some LV coefficient with dimension $1$.
\eit
Equipping with these tools and following the spirit of \cite{Adler2011bg}, we calculate the interaction energy
\bea
\hat{E}_\mathrm{int}=-\int d^3\vec{x}\mathscr{L}_\mathrm{int}\nonumber
\eea
in the following subsections.
As the analog object, we calculate the fermion-photon interaction first with the interaction Lagrangian
\bea&&\label{AIntEM}
\mathscr{L}_\mathrm{int}=\mathscr{L}_\mathrm{AI}=-g\psb\Ga^a A_a\ps\nn&&
~~~~=-g\,A_a\left[\psb\ga^a\ps-\psb\left(c_b^{~a}\ga^b+d_b^{~a}\ga_5\ga^b\right)\ps\right],
\eea
while for fermion-gravity interaction, $\mathscr{L}_\mathrm{int}$ is replaced by $\mathscr{L}_\mathrm{hI}$ in Eq. (\ref{hIntGM}).

\subsection{Non-relativistic fermion-photon interaction}
The interaction energy between two electron states $|p',\be\rangle$ and $|p,\al\rangle$ is
{\small
\bea&&
E_\mathrm{AI}\equiv E_\mathrm{AI}^\mathrm{LI}+E_\mathrm{AI}^\mathrm{LV}\nn&&
~~=g\langle p',\be|\int d^3\vec{x}\left[\psb\vec{\Ga}\ps\cdot\vec{A}-\psb\Ga^0\ps A^0\right]|p,\al\rangle\nn&&
~~=g\sum_{s_1,~s_2}\int d^3\vec{x}\int\frac{d^3k_1}{(2\pi)^3}\frac{d^3k_2}{(2\pi)^3}
\langle 0|b_\be(p')b_{s_1}^\dagger(k_1)\nn&&~~~~~~
\left[\bar{u}_{s_1}(k_1)\,\Ga\cdot A\,u_{s_2}(k_2)\right]b_{s_2}(k_2)b^\dagger_\al(p)|0\rangle
\nn&&~~
=g\int d^3\vec{x}\,e^{-i\vec{q}\cdot\vec{x}}\left[{u}^\dagger_{\be}(p')\,\ga^0\Ga\cdot A\,u_{\al}(p)\right],
\eea
}
\hspace{-1.5mm}where $q\equiv p'-p$ and $\Ga\cdot A\equiv\vec{\Ga}\cdot\vec{A}(x)-\Ga^0A^0(x)$,
and we have used $\{b_{\al}(p),~b_{\si}^\dagger(k)\}=(2\pi)^3\delta_{\al\si}\delta^3(\vec{p}-\vec{k})$.
Note we assume that the field redefinition has already been done implicitly, so for the $c,~d$ coefficients,
$\ga^0\Ga\cdot A$ has to be replaced by $(\Ga\hat{U})^\dagger\ga^0\cdot A\hat{U}$.
In the following, we will always deal with $a,~b,~H$ terms first, and treat $c,~d$ terms later,
and we omit the subscript ``$0$" for denoting LI spinor $u_{\si0}$ unless necessary.
The LI part of the interaction energy is
{\small
\bea&&\label{LIAI}
\hspace{-3mm}E_\mathrm{AI}^\mathrm{LI}=g\int d^3\vec{x}\,e^{-i\vec{q}\cdot\vec{x}}\left(\bar{u}_{\be}(p')[\vec{A}(x)\cdot\vec{\ga}-A^0(x)\ga^0]u_{\al}(p)\right)
\nn&&~~~~
\hspace{-3mm}=g\int d^3\vec{x}\,e^{-i\vec{q}\cdot\vec{x}}\xi^\dagger_{\be}\left(
\left[\frac{\vec{l}\cdot\vec{A}+i\vec{q}\times\vec{A}\cdot\vec{\si}}{E+m}\right]
    -A^0\left[1+\right.\right.
\nn&&~~~~~~\left.\left.
\frac{\vec{p'}\cdot\vec{p}+i\vec{q}\times\vec{p}\cdot\vec{\si}}{(E+m)^2}\right]
 \right)\xi_{\al}(p)\nn&&~~~~
\hspace{-3mm}=g\int d^3\vec{x}\,e^{-i\vec{q}\cdot\vec{x}}\xi_\be^\dagger\left[\left(\frac{\vec{A}\cdot\vec{p}}{m}-A^0(1+\frac{\vec{p}^2}{4m^2})\right)+
\frac{\vec{\si}\cdot\vec{B}}{2m}\right.\nn&&~~~~~~
\left.+\frac{(\vec{E}\times\vec{p})\cdot\vec{\si}}{4m^2}\right]\xi_\al,
\eea
}
\hspace{-1.5mm}which is exactly the same as the Eqs. (7.7) and (7.9) in \cite{Adler2011bg},
if the signature difference is concerned.
Note we have defined $l\equiv p'+p$, and assumed the fermion is always on the mass shell such that the energy transfer is zero, $q^0=0$ (elastic scattering),
just as in \cite{Adler2011bg}.
However, in the presence of a generic LV coefficient $X$, the dispersion relation is modified.
Thus $p'^0=p^0$ implies not $\vec{q}\cdot(2\vec{p}+\vec{q})=0$, but rather
\bea\label{deltaWX}
\frac{\vec{q}\cdot\vec{p}}{4m^2}\simeq-\frac{\vec{q}^2}{8m^2}+\frac{\delta\omega_{p'}-\delta\omega_p}{4m},
\eea
where $\delta\omega_p=\delta\omega(p,m,X)\equiv\,p^0(p,m,X)-\sqrt{\vec{p}^2+m^2}$, and we divided $\vec{q}\cdot\vec{p}$ by $4m^2$ to
fit the factor appearing in the second equation of (\ref{LIAI}).
The extra term in (\ref{deltaWX}) means that there is an extra LV contribution due to the modified dispersion relation,
even in the calculation of the superficially LI $E_\mathrm{AI}^\mathrm{LI}$.
To facilitate the analogy, we also assume that the 4-potential of the photon field is static, $\dot{A}^\mu=0$.
For simplicity, we choose the Coulomb gauge $\nabla\cdot\vec{A}=0$ , which is equivalent to the Lorenz gauge in static limit. The absence of $\vec{q}\cdot\vec{A}$ is simply due to this gauge choice.
The third term in (\ref{LIAI}) is exactly the standard Dirac's prediction, the magnetic moment interaction $\frac{\vec{\si}\cdot\vec{B}}{2m}$,
and the last term $(\vec{E}\times\vec{p})\cdot\vec{\si}$ is the spin-orbit coupling
and is responsible for the fine structure corrections.
The first term $g\vec{A}\cdot\vec{p}/m$ is simply the cross term in the gauge invariant kinetic energy
$\frac{(\vec{p}+g\vec{A})^2}{2m}$ in the Coulomb gauge, and $-gA^0$ is the static Coulomb energy with the correction factor
$1+\frac{\vec{p}^2}{4m^2}$ for a charge particle in the co-moving frame.
The vanishing of $\vec{q}^2A^0$ term is because this term is proportional to $\nabla^2A^0(\vec{x})=-\rho_e\delta(\vec{x}-\vec{x}_s)$ by Coulomb's law, where $\vec{x}_s$ denotes the position of source particle for the external EM field and the fermion is assumed to be far away from the source particle.

For LV eigenspinor contribution to EM interaction energy, first we'd like to give an explicit formula
for a generic LV coefficient $X$,
\bea&&\label{LIverLV}
\hspace{-5mm}E_\mathrm{AI}^\mathrm{X-spinor}
=g\int d^3\vec{x}\,e^{-i\vec{q}\cdot\vec{x}}\xi^\dagger_{\be}\left(
\left[\vec{\si}\cdot\vec{A}\,\delta U_X(p)+\delta U_X^\dagger(p')\,\vec{\si}\cdot\vec{A}\right]
\right.\nn&&~~
\left.-A^0\left[U_0^\dagger(p')\delta U_X(p)+\delta U_X^\dagger(p') U_0(p) \right]
\right)
\xi_{\al}(p),
\eea
where $U_0(p)\equiv\frac{\vec{\si}\cdot\vec{p}}{\omega_0+m}$ and $U_X$ are the LI and LV matrices
connecting the upper and lower Pauli spinors. For example, for a given Dirac spinor
$u(p)=\left(
        \begin{array}{cc}
          \xi(p), & \eta(p) \\
        \end{array}
      \right)^\mathrm{T}$, $\eta=U_X\xi$ and $\delta U_X(p)\equiv\,U_X(p)-U_0(p)$.
For details, see Appendix \ref{LVeigenSpinors}.

First, we can calculate the contribution of $a,~b,~H$ coefficients to the fermion-photon interaction separately.
As they do not superficially alter the conserved currents,
there is no modification of fermion-photon vertex due to these coefficients.
In other words, for $a,~b,~H$, it is sufficient to take into account of eigen-spinor contributions $E_\mathrm{AI}^\mathrm{X-spinor}$ and corrections due to modified dispersion relations.
For the $a$-coefficient, its effects can be simply shown by replacing
$\vec{p}\rightarrow\vec{p}+\vec{a}$ in (\ref{LIAI}) and omitting $\vec{a}^2$ terms, which are of higher order.
This manifestly shows that for EM interaction, $a$-term only shift the 4-momentum
and cause no observable physical effects, and thus can be removed by proper field redefinitions \cite{SMEa}\cite{TassonMGC}.
However, it does have effects for gravitational interaction \cite{TassonMGC}, and will be explicitly shown in the next subsection.

The LV correction for $b$-coefficient is
\bea\label{LVAI-b}&&
\hspace{-5mm}E_\mathrm{AI}^b=g\int d^3\vec{x}\,e^{-i\vec{q}\cdot\vec{x}}\xi_\be^\dagger\left[\frac{i(\vec{q}\times\vec{A})\cdot\vec{b}
-b^0A^0\vec{l}\cdot\vec{\si}/2}{2m^2}\right.\nn&&
\hspace{-2mm}\left.+\frac{b^0\vec{\si}\cdot\vec{A}}{m}+\frac{\vec{A}\cdot(\vec{l}\vec{b}-\vec{b}\vec{l})
\cdot\vec{\si}}{2m^2})\right]\xi_\al\nn&&
\hspace{-3mm}=g\int d^3\vec{x}\,e^{-i\vec{q}\cdot\vec{x}}\xi_\be^\dagger\left[\frac{b^0\vec{\si}\cdot\vec{A}}{m}+\frac{\vec{B}\cdot\vec{b}
-b^0A^0\vec{p}\cdot\vec{\si}}{2m^2}\right.\nn&&
\hspace{-2mm}\left.+\frac{2\left[\vec{A}\cdot\vec{p}\,(\vec{\si}\cdot\vec{b})-\vec{A}\cdot\vec{b}\,(\vec{\si}\cdot\vec{p})\right]
+i\vec{\si}\cdot\vec{\nabla}(\vec{b}\cdot\vec{A})}{2m^2})\right]\xi_\al.
\eea
Note that $-\frac{b^0A^0\vec{q}\cdot\vec{\si}}{4m^2}$ in (\ref{LVAI-b}) is canceled by the correction due to LV dispersion relation,
see Eq. (\ref{deltaWX}).
Interestingly, the $\vec{B}\cdot\vec{b}/2m^2$ term seems indicate that the $\vec{b}$ vector behaves like a ``cosmic magnetic dipole moment" $\delta\vec{\mu}=-\frac{g\vec{b}}{2m^2}$ in comparison to the conventional magnetic dipole moment (MDM)  $\vec{\mu}=-\frac{g\vec{\si}}{2m}$.
Contrary to the dynamical $\vec{\mu}$, which can be manipulated by spin polarization,
$\delta\vec{\mu}$ is supposed to be a constant background, whose projection on a specific direction, say $\vec{B}$,
varies due to the relative motion of the charged particle with respect to the cosmic backgrounds,
and may cause a sidereal variation in terrestrial experiments.
A similar $\vec{\Omega}\cdot\vec{b}$ coupling also arises when a LV fermion is coupled to the gravitational field due to a large rotating mass.
For a fermion coupled with some kind of cosmic anisotropic vector \cite{WTCASc}, or the axial vector part of a torsion tensor
by the identification $\frac{\epsilon^{\mu\al\be\ga}}{8}T_{\al\be\ga}\rightarrow b_\mathrm{eff}^\mu$ \cite{TassonMGC},
we may expect similar forms of interaction.
In fact, the non-minimal $\hf b_F^{ijk}F_{jk}\psb\ga_5\ga_i\ps$ term \cite{Ding2016} may also produce a term looking like $b_F\vec{\si}\cdot\vec{B}$ with similar structure if $b_F^{ijk}=b_F\epsilon^{ijk}$, and thus the terms within quite different scenarios may be constrained by similar phenomenological observations, such as the comagnetometer experiments \cite{comagnet1}\cite{comagnet2}.

The LV correction for $H$-coefficient is
\bea\label{LVAI-H}&&
\hspace{-2mm}E_\mathrm{AI}^H
=g\int d^3\vec{x}\,e^{-i\vec{q}\cdot\vec{x}}\xi_\be^\dagger\left[\frac{\vec{A}\times\vec{H}\cdot\vec{\si}}{m}
-\frac{(\vec{\si}\cdot\vec{A})\,(\vec{l}\cdot\vec{\tilde{H}})}{2m^2}\right.\nn&&~~
\left.+\frac{iA^0\vec{q}\cdot\vec{H}}{4m^2}
-\frac{A^0\vec{l}\times\vec{H}\cdot\vec{\si}}{4m^2}\right]\xi_\al\nn&&~~
=g\int d^3\vec{x}\,e^{-i\vec{q}\cdot\vec{x}}\xi_\be^\dagger\left[\frac{\vec{A}\times\vec{H}\cdot\vec{\si}}{m}
-\frac{(\vec{\si}\cdot\vec{A})\,(\vec{p}+\vec{q}/2)\cdot\vec{\tilde{H}}}{m^2}\right.\nn&&~~
\left.
-\frac{\vec{E}\cdot\vec{H}}{4m^2}-\frac{A^0(\vec{p}+\vec{q})\times\vec{H}\cdot\vec{\si}}{2m^2}\right]\xi_\al,
\eea
where we have decomposed $H_{\mn}$ into an ``electric" part $\vec{H}^i\equiv H_{0i}$ and a ``magnetic" part
$\vec{\tilde{H}}^i\equiv\hf\epsilon_{ijk}H_{jk}$.
This decomposition is meaningful, as seen from various couplings such as $-\vec{E}\cdot\vec{H}/4m^2$.
Just like the ``cosmic MDM" induced by the $\vec{b}$ vector, $-\vec{E}\cdot\vec{H}/4m^2$ behaves
like a ``cosmic electric dipole moment" $\frac{g\vec{H}}{4m^2}$ for a charged fermion.
Also like $-b^0A^0\vec{\si}\cdot\vec{p}/2m^2$ in (\ref{LVAI-b}), $A^0\vec{H}\times\vec{p}\cdot\vec{\si}/2m^2=-A^0\vec{\si}\times\vec{p}\cdot\vec{H}/2m^2$ induces a tiny LV spin-orbit (SO) correction to the LI counter term $g\frac{(\vec{E}\times\vec{p})\cdot\vec{\si}}{4m^2}$.
However, the external $\vec{E}$ in the LI operator is controllable, while the ``cosmic" $\vec{H}$ term is not,
though it may receives a sidereal variation for any terrestrial experiment.
Moreover, it depends on the local electric potential, which is like the term $\phi_g\frac{\vec{p}\times\vec{H}\cdot\vec{\si}}{m}$ in (\ref{GI-H1}).
These distinctive features means the LV SO couplings can be testable and distinguished from any LI background
in the ultrahigh precision fine structure observations.

For the $c,~d$-coefficients, as they not only lead to eigen-spinor corrections,
but also bring corrections to conserved current, and hence impose the need of
spinor redefinition to cure the otherwise non-hermitian Hamiltonian if the
spinor $\psi$ is improperly used, we treat them separately.

After redefinition, the fermion-photon interaction is
\bea\label{FPI-cd}&&
\hspace{-6mm}gA_a\psb[\delta_b^{~a}-c_b^{~a}-d_b^{~a}\ga_5]\ga^b\ps
=g\chi^\dagger\left[(\vec{\al}\cdot\vec{A}-A^0)\right.\nn&&
\hspace{-4mm}\left.+{A}^j(\tilde{d}_{ij}\ga_5\al^i+2d_{(0j)}\ga_5)-{A}^j(\tilde{c}_{ij}\al^i+2c_{(0j)})\right]\chi,
\eea
where we defined $c_{(0j)}\equiv\hf(c_{0j}+c_{j0})$, $d_{(0j)}\equiv\hf(d_{0j}+d_{j0})$,
$\tilde{c}_{ij}\equiv\,c_{00}\delta_{ij}+c_{ij}$,~$\tilde{d}_{ij}\equiv\,d_{00}\delta_{ij}+d_{ij}$,
and again we preserve terms only up to linear order of LV coefficients.
The LV $c,~d$ corrections to the conserved current are the terms in the second line in (\ref{FPI-cd}), where
the terms in the first line in the large bracket correspond to LI current.
It is interesting to note that the consistency of the field redefinition for $c,~d$ terms lies in the fact that
there is no LV $A^0$ coupling operator in the second line in (\ref{FPI-cd}), as the goal of the field redefinition
is just to remove the unconventional kinematic couplings caused by the $c,~d$ terms, and the $A^0$ coupling will in no doubt
be removed due to {\it the minimal coupling} schemes.
Inserting the quantization expansion of $\chi$ in terms of annihilation and creation operators as Eq. (\ref{fermion}), we get
\bea\label{AICur-c}&&
\hspace{-6mm}E_\mathrm{AI}^\mathrm{c_1}=-g\int d^3\vec{x}\,e^{-i\vec{q}\cdot\vec{x}}\,
u^\dagger_{\be}(p')\left[\al^i\tilde{c}_{ij}A^j+2c_{(0j)}A^j\right]u_{\al}(p)\nn&&
\hspace{-2mm}=-g\int\,d^3\vec{x}\,e^{-i\vec{q}\cdot\vec{x}}\,\xi^\dagger_\be\,\left[2\vec{c}\cdot\vec{A}[1+\frac{\vec{p'}\cdot\vec{p}
+i\vec{q}\times\vec{p}\cdot\vec{\si}}{4m^2}]\right.\nn&&
\hspace{-2mm}\left.+\frac{c_{00}}{2m}[\vec{l}\cdot\vec{A}+\vec{B}\cdot\vec{\si}]+\frac{c_{ij}A^j}{2m}(l^i+i\epsilon_{kil}q^k\si^l)\right]\xi_\al,
\\\label{AICur-d}&&
\hspace{-6mm}E_\mathrm{AI}^\mathrm{d_1}=
g\int\,d^3\vec{x}\,e^{-i\vec{q}\cdot\vec{x}}\,u^\dagger_{\be}(p')\left\{\tilde{d}_{ij}\Si^iA^j
+2\vec{d}\cdot\vec{A}\ga_5\right\}u_{\al}(p)\nn&&
\hspace{-2mm}=g\int\,d^3\vec{x}\,e^{-i\vec{q}\cdot\vec{x}}\,\xi^\dagger_\be\,\left\{\vec{d}\cdot\vec{A}\frac{\vec{\si}\cdot\vec{l}}{m}
+\tilde{d}_{ij}A^j\cdot\left[\si^i+\right.\right.\nn&&
\hspace{-2.mm}\left.\left.\frac{(2p^i+q^i)\,\vec{\si}\cdot\vec{p}+p^i\,\vec{\si}\cdot\vec{q}
-\vec{p'}\cdot\vec{p}\,\si^i+i(\vec{p}\times\vec{q})^i}{4m^2}\right]\right\}\xi_\al,
\eea
where we defined $\vec{c}^j\equiv c_{(0j)}$ and $\vec{d}^j\equiv d_{(0j)}$ for notational simplicity.
Note both in (\ref{AICur-c}) and (\ref{AICur-d}), $u_{\si}(k)=u_{\si0}(\vec{k})$ in (\ref{LIES}),
as we only need to keep terms of linear order of LV coefficients.
By the comparison of (\ref{AICur-c}) with (\ref{LIAI}), we see $c_{00}$ acts like a scale factor to the LI
counter terms such as $\frac{\vec{\si}\cdot\vec{B}}{m}$,~$\frac{\vec{p}\cdot\vec{A}}{m}$, while
$c_{ij}$ plays the role of a shear factor, and $\vec{c}^i=c_{(0i)}$ mixes the coupling of $\vec{A}$ into those
originally coupled with $A^0$ if LV is absent.
In short, $c_{\mn}$ acts like a metric tensor:
it not only scale isotropically, but also shear slightly the original LI EM interactions, as if
the original terms being viewed in a slightly sheared coordinates.
However, we should avoid the confusion with the so-called ``passive coordinate transformations", which have no
physical effects \cite{SMEa}. In comparison, the $c$-coefficient induced effects are in principle testable, such as constraints of
the sidereal variation by measuring the transition frequency in atomic clocks \cite{NatSanner2018}.
For $d$-coefficient, due to the $\ga_5$ factor, it mediates the spin-orbit couplings with the EM field,
except the $\tilde{d}_{ij}A^j\si^i$ and $\frac{i\tilde{d}_{ij}A^j(\vec{p}\times\vec{q})^i}{4m^2}$ terms.
For example, the $\frac{\tilde{d}_{ij}A^jp^i\vec{\si}\cdot\vec{p}}{2m^2}$ term looks much like an anomalous magnetic moment (AMM) coupling term
$\frac{\mu'(\vec{B}\cdot\vec{p})(\vec{\si}\cdot\vec{p})}{2m^2}$ \cite{Silenko2003-FW}, where $\mu'$ is the AMM coupling constant put by hand.

Next we consider the LV eigen-spinor corrections to the superficially LI term, the term in the first line in the
large square bracket in (\ref{FPI-cd}).
The eigen-spinor for $c,~d$-coefficients in the quantization of $\chi$ has to be obtained
from the free LV modified Dirac equation,
\bea&&\label{cdFDirac}
i\dot{\chi}=-i\left[(\delta_{ij}-\tilde{c}_{ij}+\tilde{d}_{ij}\ga_5)\al^i-2(c_{(0j)}-d_{(0j)}\ga_5)\right]\nabla_j\chi\nn&&~~~~
+m\left[\ga^0(1-c_{00})-d_{j0}\ga_5\ga^j\right]\chi.
\hspace{-5mm}
\eea
Assuming the eigen-spinor takes the form
$\chi=e^{ip\cdot x}
\left(
\begin{array}{ccc}
  \xi    \\
  \eta   \\
\end{array}
\right)$, where $\eta=U_X\xi$, we obtain the $U_X$ with $X=c,~d$, see (\ref{UdMc},\ref{UcMc}) in the Appendix.

We still treat $c,~d$ terms separately in the spirit of keeping only linear order of LV coefficients.
For the $c$-coefficient, the LV eigen-spinor contrition to EM interaction can be obtained by substituting
$\delta U_c(k)\equiv U_c(k)-U_0(k)$ in (\ref{LIverLV}),
\bea\label{AIES-c2}&&
\hspace{-3mm}E_\mathrm{AI}^\mathrm{c_2}=g\int d^3\vec{x}\,e^{-i\vec{q}\cdot\vec{x}}\xi^\dagger_{\be}\left\{                                                                   \left[A^0(-\frac{c_{(0j)}q^j}{2m}+\frac{c_{ij}p^ip^j}{2m^2})\right.\right.
\nn&&~~\left.\left.
-c_{ij}\frac{p^jA^i}{m}\right]+\left[A^0\frac{2ic_{ij}q^{[j}p^{k]}\epsilon_{ikl}\si^l-c_{ij}q^iq^j}{4m^2}
\right.\right.
\nn&&~~\left.\left.
-c_{ij}\frac{q^jA^i+i\epsilon_{ikl}q^jA^k\si^l}{2m}\right]\right\}
\xi_{\al}(p),
\eea
again we have added the correction $A^0(\delta\omega_p'-\delta\omega_p)/{4m}$ into (\ref{AIES-c2})
by substituting (\ref{deltac}), see Eq.(\ref{deltaWX}).
The total LV fermion-photon interaction energy due to $c$-coefficient is $E_\mathrm{AI}^\mathrm{c_1}+E_\mathrm{AI}^\mathrm{c_2}$.

For the $d$-coefficients, the LV eigen-spinor correction is
\bea\label{AIES-d2}&&
\hspace{-5mm}E_\mathrm{AI}^\mathrm{d_2}=g\int d^3\vec{x}\,e^{-i\vec{q}\cdot\vec{x}}\xi^\dagger_{\be}\left(                                                                   A^0\left[\frac{d_{j0}(q^{(i}p^{j)}+\frac{q^jq^i}{2})+d_{0j}q^iq^j}{4m^2}\right.\right.\nn&&~~
\hspace{-3mm}\left.\left.+\frac{\tilde{d}_{ij}q^j}{4m}-\frac{d_{0j}p^ip^j}{2m^2}\right]\si^i
-\frac{i\epsilon_{jkl}d_{ji}A^l(2q^{(k}p^{i)}+q^kq^i)}{2m^2}
\right.\nn&&~~
\hspace{-3mm}\left.
+d_{ji}\frac{2A^{[j}\si^{k]}[p^kp^i+q^{(k}p^{i)}+q^kq^i/2]}{m^2}+d_{0j}\frac{l^j\,\vec{\si}\cdot\vec{A}}{2m}\right)
\xi_{\al}(p).\nn
\eea
Unlike (\ref{AICur-c}) and (\ref{AICur-d}), $E_\mathrm{AI}^\mathrm{c_2}$ and $E_\mathrm{AI}^\mathrm{d_2}$ do contain contributions from the LV interaction with scalar potential $A^0$.
These terms would be absent if corrections from LV eigen-spinors were not taken into account, see the second lines in (\ref{FPI-cd}).
The total LV fermion-photon interaction energy due to $d$-coefficient is $E_\mathrm{AI}^\mathrm{d_1}+E_\mathrm{AI}^\mathrm{d_2}$.
We separately write them out in order to make the nature of where they originate (from LV corrected current or LV eigen-spinor) more clear.
Also note, the vector potential $\vec{A}$ replaces the scalar potential $A^0$ in the SO coupling term $c_{ij}\frac{E^j(\vec{\si}\times\vec{p})^i-p^j(\vec{\si}\times\vec{E})^i}{4m^2}$ in (\ref{AIES-c2}),
for the corresponding $d_{ij}$ term $-\frac{i\epsilon_{jkl}d_{ji}A^lq^{(k}p^{i)}}{m^2}$ in (\ref{AIES-d2}).
The reason can date back to the additional $\ga_5$ factor in LV kinematic $d$-term compared with the corresponding $c$-term in $\mathcal{L}\supset\frac{i}{2}\psb\,\delta\Ga^\mu\lrprt\mu\psi$.

\subsection{Non-relativistic fermion-gravity interaction}\label{FGI-NR}
Now we calculate the fermion-gravity interaction.
Since we assume the scattered fermions are on the mass shell, which means the equation of motion is satisfied,
the term proportional to $h$ does not contribute.
The interaction energy $-\hf\int\,d^3x\,h_{\mn}T^{\mn}$ in (\ref{hIntGM}) is proportional to
\bea&&\label{hIntGMS}
\hf h^b_{~a}\left(\frac{i}{2}\psb\Ga^a\lrdrt b\ps-\psb\left[(a_b+b_b\ga_5)\ga^a
+H_{bc}\si^{ac}\right]\ps\right)\nn&&~~
-\frac{i}{4}\bar{\psi}\left[h^{\nu a}(c_{b\nu}+d_{b\nu}\ga_5)
+h^{\rho}_{~b}(c_\rho^{~a}+d_\rho^{~a}\ga_5)\right]\ga^b\lrdrt a\ps
\nn&&~~
-\frac{1}{4}\epsilon^{bcmn}h_{am,n}\psb[c_b^{~a}\ga_5+d_b^{~a}]\ga_c\ps.
\eea
Note we ignore all photon couplings by replacing $D_\mu\rightarrow\prt_\mu$, not only for calculational
simplicity, but also to facilitate the discussions of the test of equivalence principle (EP),
where photon interaction not only complicates, but may even spoil the precision test of weak EP \cite{Drummond1980-EP}.

The LI fermion-gravity interaction is
{\small
\bea\label{LIGI0}&&
\hspace{-5mm}E_\mathrm{GI}^\mathrm{LI}=
{_0\langle} p',\be|\int\,d^3x\,\left\{\frac{i}{4}h^b_{~a}\psb\ga^a\lrprt b\ps\right\}|p,\al\rangle_0\nn&&
\hspace{-5mm}=
\int\,d^3x\,e^{-i\vec{q}\cdot\vec{x}}\xi^\dagger_{\be}\left\{
p^0\left[\phi_g(1+\frac{\vec{p}^2+\vec{q}\cdot\vec{p}+i\vec{q}\times\vec{p}\cdot\vec{\si}}{4m^2})\right.\right.\nn&&
\left.\left.-\frac{\vec{A}_g\cdot\vec{l}+i\vec{q}\times\vec{A}_g\cdot\vec{\si}}{4m}\right]
+(\vec{p}+\frac{\vec{q}}{2})\cdot\left[\frac{\phi_g}{2m}(\vec{l}-i\vec{q}\times\vec{\si})\right.\right.\nn&&
\left.\left.
-\frac{\vec{A}_g}{2}(1+\frac{\vec{p}^2+\vec{q}\cdot\vec{p}+i\vec{q}\times\vec{p}\cdot\vec{\si}}{4m^2})\right]
\right\}\xi_{\al}\nn&&
\hspace{-5mm}=\int\,d^3x\,e^{-i\vec{q}\cdot\vec{x}}\xi^\dagger_{\be}\left\{m\phi_g(1+\frac{7\vec{p}^2}{4m^2})
+\frac{3\vec{g}\times\vec{p}\cdot\vec{\si}}{4m}
-\vec{A_g}\cdot\vec{p}\right.\nn&&
\left.-\frac{\vec{\Omega}\cdot\vec{\si}}{4}\right\}\xi_{\al},
\eea}\hspace{-1.5mm}
where $|p,\al\rangle_0$ denotes the pure LI eigen-vector, and we utilized the conventional definition of the GEM vector potential, which differs slight from the definition of $A_g^j$ in the Appendix \ref{GraviEM}, \ie,
\bea\label{WeakGraviP}&&
h_{00}=-2\phi_g,\quad h_{ij}=-\delta_{ij}2\phi_g,\quad h_{0j}=h_{j0}=A_g^j,\nn&&
\vec{g}\equiv\nabla\phi_g,\quad  \vec{\Omega}\equiv\nabla\times\vec{A}_g.
\eea
We also preserve only terms up to the first Post-Newtonian order, \ie, PNO(1).
Specific in detail, we keep
\bea\label{orderE}
h_{00}\sim\mathcal{O}(v^{4}),\quad h_{0i}\sim\mathcal{O}(v^{3}),\quad h_{ij}\sim\mathcal{O}(v^{2}).
\eea
Though the metric component $h_{00}=-2\phi_g$ only contains $\mathcal{O}(v^{2})$ term $\phi_g$, it is compensated by keeping the NR factors, such as $\phi_g\frac{\vec{p}^2}{m^2}\sim\mathcal{O}(v^{4})$.
Also note only a quarter of the spin-orbit FG interaction $\frac{3\vec{g}\times\vec{p}\cdot\vec{\si}}{4m}$
comes from the pure temporal metric $h_{00}$ contribution, and the other half comes from the spatial metric $h_{ij}$
contribution.
In comparison with the Eq. (20) in \cite{WTHehl1990},
the spin-orbit operator from $h_{00}$ exactly coincides with the corresponding term due to
a non-relativistic fermion coupled with the non-inertial force, thus confirms the weak EP \cite{WTHehl1990},
since the only nonzero metric perturbation for a linear acceleration is $h_{00}=\vec{a}\cdot\vec{r}$.

It will be interesting to compare the result (\ref{LIGI0}) with those obtained by the FW
transformation, such as the Eq.(2.44) in \cite{Fischbach} for a static spherically symmetric metric.
For that purpose, we also keep
\bea\label{orderEe}&&
\prt_ih_{00}|\frac{p^i}{m}|\sim\frac{m\lambda_c}{\bar{r}}\mathcal{O}(v^{3}),~~ h_{0i}p^k\sim m\mathcal{O}(v^{4}),\nn&&
\prt_jh_{0i}\sim\frac{m\lambda_c}{\bar{r}}\mathcal{O}(v^{3}),
\eea
where $\la_c=\frac{\hbar}{mc}$ is the Compton wavelength of the fermion we concerned, say, a neutron,
and $\bar{r}$ is the characteristic length scale of the gravitational source, such as the Earth radius.
In general, $\la_c/\bar{r}\ll\frac{v}{c}$ (we temporarily restore $c$ for clearness), for \eg, on the Earth, $\la_c/\bar{r}\sim10^{-22}$ and $v/c\sim10^{-6}$ for thermal neutron ($v\sim10^3\mathrm{m}/\mathrm{s}$),
thus on numerical grounds, we can totally ignore terms involving $\prt_jh_{0i}$ and $\prt_ih_{00}|\frac{\vec{p}}{m}|$.
However, not only for parallel comparison, but also in preparation for exotic situations such as neutron stars,
where both $h_{00}\sim0.1,~|\vec{\Omega}|\sim10^{-23}$GeV are much larger than the
corresponding values on Earth, we keep these terms in the following.
In deriving Eq.(\ref{LIGI0}), we also assume $\vec{\Omega}$ is constant such that $\vec{A}_g=\hf\vec{\Omega}\times\vec{r}$,
and utilize the equations $\nabla^2\phi_g=4\pi\,G_N\rho_m$ and $\nabla^2\vec{A}_g=16\pi\,G_N\vec{j}_m$
to eliminate the $\vec{p}\cdot\vec{q}\,\phi_g$ and $\vec{p}\cdot\vec{q}\,\vec{A}_g$ terms,
since the neutral fermion is assumed to be outside the matter source of gravity, where $\rho_m$ and $\vec{j}_m$ vanish.
This also explains why there is no $i\vec{g}\cdot\vec{p}$ term compared with the NR Hamiltonian obtained by the FW approach.

Now we also consider the $a,~b,~H$ terms first as these terms do not involve derivative couplings.
However, unlike the EM current $\psb\Ga^a\ps$, the $a,~b,~H$ terms do contribute to the LV energy-momentum tensor $T^{\mn}$, which receives any kind of contribution from matter source.
For simplicity, we discuss the LV eigen-spinor corrections first.
As in Eq.(\ref{LIverLV}),
we write down a general formula for the LV eigen-spinor correction,
{\small\bea\label{GFSpinLV}&&
E_\mathrm{GI}^\mathrm{X-spinor}
=\frac{1}{4}\int\,d^3x\,e^{-i\vec{q}\cdot\vec{x}}\xi^\dagger_\be\left\{
[2l^0\phi_g-\vec{l}\cdot\vec{A}_g]\right.
\nn&&~~~~~~\left.
\cdot\left[\delta{U}_X^\dagger(\vec{p'})\,U_0(\vec{p})+
U_0^\dagger(\vec{p'})\,\delta{U}_X(\vec{p})\right]+[2\phi_g\vec{l}-\vec{A}_gl^0]
\right.
\nn&&~~~~~~\left.
\cdot\left[\delta{U}_X^\dagger(\vec{p'})\,\vec{\si}
+\vec{\si}\,\delta{U}_X(\vec{p})\right]\right\}\xi_\al,
\eea
}
\hspace{-1.5mm}where the $2\times2$ matrices $U_0$ and $\delta{U}_X(\vec{p})={U}_X-U_0$ are defined in (\ref{LIverLV}).
Note the structure similarity between (\ref{LIverLV}) and (\ref{GFSpinLV}), where $\phi_g$ and $A_g$ replace the role
of $\phi$ and $\vec{A}$, respectively,
while the remain terms, $\vec{A}_g\cdot\vec{l}$ and $\phi_g\vec{l}$, reflect the tensor nature of gravitational coupling.
Both the similarity and difference between FG and FE couplings may stem from this peculiar structure.

Substituting $\delta{U}_X(\vec{p})$ with $X=a,~b,~H,~c,~d\,$ separarely into (\ref{GFSpinLV}) with $\delta{U}_X$
given in Appendix \ref{LVeigenSpinors}, the LV eigen-spinor corrections to FG interaction
due to $a,~b,~H$ coefficients are
{\small\bea\label{GI-ab1}&&
\hspace{-3mm}E_\mathrm{GI}^{ab-1}=\frac{1}{4}\int\,d^3x\,e^{-i\vec{q}\cdot\vec{x}}\xi^\dagger_\be\left\{
[2l^0\phi_g-\vec{l}\cdot\vec{A}_g]\left[\frac{(\vec{a}+b^0\vec{\si})\cdot\vec{l}}{4m^2}\right.\right.\nn&&~~
\left.\left.+\frac{i\vec{q}\times\vec{a}\cdot\vec{\si}}{4m^2}\right]
+[2\phi_g\vec{l}-\vec{A}_gl^0]\cdot\left[\frac{\vec{a}+b^0\vec{\si}}{m}-
\frac{i\vec{q}\times\vec{b}}{2m^2}-\right.\right.\nn&&~~
\left.\left.
\frac{(\vec{l}\times\vec{b})\times\vec{\si}}{2m^2}\right]\right\}\xi_\al
\nn&&\simeq
\int\,d^3x\,e^{-i\vec{q}\cdot\vec{x}}\xi^\dagger_\be\left\{
\frac{\vec{g}\times\vec{a}\cdot\vec{\si}}{4m}+
\phi_g\left[\frac{3\vec{a}\cdot(\vec{p}+\vec{q}/2)}{2m}+\right.\right.\nn&&
~~\left.\left.
\frac{3b^0\vec{l}\cdot\vec{\si}}{4m}+\frac{(\vec{b}\cdot\vec{l})(\vec{\si}\cdot\vec{l})}{4m^2}
-\frac{(\vec{p}^2+\vec{p}\times\vec{q}+\vec{q}^2/4)}{m^2}(\vec{b}\cdot\vec{\si})\right]
\right.\nn&&
~~\left.-\left[\vec{A}_g\cdot(\vec{a}+\frac{b^0\vec{\si}}{2})+\frac{\vec{\Omega}\cdot\vec{b}}{4m}
+\frac{\vec{A}_g\cdot(\vec{l}\vec{b}-\vec{b}\vec{l})\cdot\vec{\si}}{4m}\right]\right\}\xi_\al,\\
\label{GI-H1}&&
\hspace{-3mm}E_\mathrm{GI}^{H-1}=\frac{1}{4}\int\,d^3x\,e^{-i\vec{q}\cdot\vec{x}}\xi^\dagger_\be
\left\{[2l^0\phi_g-\vec{l}\cdot\vec{A}_g]\left[\frac{\vec{l}\times\vec{H}\cdot\vec{\si}}{4m^2}\right.\right.\nn&&~~
\left.\left.
-\frac{i\vec{q}\cdot\vec{H}}{4m^2}\right]
+[2\phi_g\vec{l}-\vec{A}_gl^0]\cdot\left[\frac{(\vec{H}\times\vec{\si})}{m}
-\frac{\vec{l}\cdot\vec{\tilde{H}}}{2m^2}\vec{\si}\right]\right\}\xi_\al\nn&&
\simeq
\int\,d^3x\,e^{-i\vec{q}\cdot\vec{x}}\xi^\dagger_\be\left\{\phi_g
\left[\frac{3\vec{l}\times\vec{H}\cdot\vec{\si}}{4m}-\vec{l}\cdot\vec{\si}\frac{\vec{l}\cdot\vec{\tilde{H}}}{4m^2}\right]
-\frac{\vec{g}\cdot\vec{H}}{4m}
\right.\nn&&~~\left.
-\vec{A}_g\cdot\frac{(\vec{H}\times\vec{\si})}{2}
+\vec{A}_g\cdot\vec{\si}\frac{\vec{l}\cdot\vec{\tilde{H}}}{4m}
\right\}\xi_\al,
\eea
}\hspace{-1.5mm}
where we have ignored terms of order $\frac{\vec{g}}{m^2}$ and $\frac{\vec{A}_g}{m^2}$ in the last two approximations.
Note $\frac{\vec{\Omega}\cdot\vec{b}}{4m}$ appears in (\ref{GI-ab1}) just as $\frac{g\vec{B}\cdot\vec{b}}{2m^2}$ appears in (\ref{LVAI-b}).
The less suppression by the inverse power of $m$ is due to the fact that in the gravitational case $m$ plays the role of coupling constant $g$.
As mentioned before, the parallel terms can be found from (\ref{LVAI-b}) and (\ref{LVAI-H}) by replacing $\phi,\vec{A}$ with $\phi_g,\vec{A}_g$,
though the associated numerical factors are different. Due to the tensor nature of gravity, there are addition terms such as $-\phi_g\vec{l}\cdot\vec{\si}\frac{\vec{l}\cdot\vec{\tilde{H}}}{4m^2}$, $\phi_g\frac{(\vec{b}\cdot\vec{l})(\vec{\si}\cdot\vec{l})}{4m^2}$ and $\frac{-\vec{p}^2\phi_g\vec{b}\cdot\vec{\si}}{4m^2}$ in comparison with the FE couplings for
the corresponding LV coefficients.

For the apparent LV interaction vertices due to $a,~b,~H$, their contribution to the interaction energy is
{\small\bea\label{GFab2}&&
\hspace{-5mm}E_\mathrm{GI}^\mathrm{ab-2}=
-\hf\langle p',\be|\int\,d^3x\,\left\{h_{ba}\psb\left[(a^b+b^b\ga_5)\ga^a\right]\ps\right\}|p,\al\rangle\nn&&
\hspace{-3mm}=\int\,d^3x\,e^{-i\vec{q}\cdot\vec{x}}\xi_\be^\dagger\left\{(\phi_{g}a^{0}-\frac{\vec{a}\cdot\vec{A}_{g}}{2})
(1+\frac{\vec{p}'\cdot\vec{p}+i\vec{q}\times\vec{p}\cdot\vec{\si}}{4m^{2}})\right.
\nn&&
\hspace{-3mm}\left.+[\frac{\phi_{g}\vec{a}}{2m}-\frac{\vec{A}_{g}a^0}{4m}]\cdot[\vec{l}-i\vec{q}\times\vec{\si}]+
(\frac{\vec{A}_{g}\cdot\vec{b}}{2}-\phi_{g}b^{0})\frac{\vec{\sigma}\cdot\vec{l}}{2m}\right.
\nn&&
\hspace{-3mm}\left.
+(\frac{b^{0}\vec{A}_{g}}{2}-\phi_{g}\vec{b})
\cdot\left[(1-\frac{\vec{p}\cdot\vec{p'}}{4m^2})\vec{\sigma}+
\frac{i\vec{p}\times\vec{q}+(\vec{p}\vec{p'}+\vec{p'}\vec{p})\cdot\vec{\si}}{4m^2}\right]\right\}\xi_\al\nn&&
\hspace{-3mm}\simeq\int\,d^3x\,e^{-i\vec{q}\cdot\vec{x}}\xi_\be^\dagger\left\{\phi_{g}\left[a^{0}(1+\frac{\vec{p}'\cdot\vec{p}}{4m^{2}})
+\frac{(\vec{a}-b^{0}\vec{\sigma})\cdot\vec{l}}{2m}\right]-\frac{\vec{a}\cdot\vec{A}_{g}}{2}\right.
\nn&&
\hspace{-3mm}\left.
+\frac{\vec{g}\times\vec{a}\cdot\vec{\si}}{2m}-a^0\left[\frac{\vec{A}_{g}\cdot\vec{l}+\vec{\Omega}\cdot\vec{\si}}{4m}\right]+
\frac{\vec{A}_{g}}{2}\cdot(b^{0}\vec{\si}+\frac{\vec{b}\,\vec{\sigma}\cdot\vec{l}}{2m})\right.
\nn&&
\hspace{-3mm}\left.
-\phi_{g}\vec{b}\cdot\left[(1-\frac{\vec{p}\cdot\vec{p'}}{4m^2})\vec{\sigma}
+\frac{(\vec{p}\vec{p'}+\vec{p'}\vec{p})\cdot\vec{\si}}{4m^2}\right]
\right\}\xi_\al.
\eea
\bea\label{GFH2}&&
\hspace{-5mm}E_\mathrm{GI}^\mathrm{H-2}=
-\hf\langle p',\be|\int\,d^3x\,\left\{h_{ba}\psb\,
H^{bc}\si^a_{~c}\ps\right\}|p,\al\rangle\nn&&
\hspace{-3mm}=\int\,d^3x\,e^{-i\vec{q}\cdot\vec{x}}\xi_\be^\dagger\left\{
(\frac{\vec{A}_g\times\vec{H}}{2}+2\phi_g\vec{\tilde{H}})\cdot
\left[(1+\frac{\vec{p}\cdot\vec{p'}}{4m^2})\vec{\sigma}-\right.\right.
\nn&&\left.\left.
\frac{i\vec{p}\times\vec{q}+(\vec{p}\vec{p'}+\vec{p'}\vec{p})\cdot\vec{\si}}{4m^2}\right]
+\frac{\vec{\tilde{H}}}{4m}\cdot\left(\vec{l}\vec{\si}-\vec{\si}\vec{l}\right)\cdot\vec{A}_g
\right.
\nn&&\left.
-\frac{i\vec{q}\times\vec{A}_g\cdot\vec{\tilde{H}}}{4m}
\right\}\xi_\al\nn&&
\hspace{-3mm}\simeq\int\,d^3x\,e^{-i\vec{q}\cdot\vec{x}}\xi_\be^\dagger\left\{\frac{\vec{A}_g\times\vec{H}}{2}\cdot\vec{\si}
+\frac{\vec{\tilde{H}}}{4m}\cdot\left[(\vec{l}\vec{\si}-\vec{\si}\vec{l})\cdot\vec{A}_g\right]
\right.
\nn&&\left.
+2\phi_g\vec{\tilde{H}}\cdot
\left[(1+\frac{\vec{p}\cdot\vec{p'}}{4m^2})\vec{\sigma}+\frac{(\vec{p}\vec{p'}+\vec{p'}\vec{p})\cdot\vec{\si}}{4m^2}\right]
-\frac{\vec{\Omega}\cdot\vec{\tilde{H}}}{4m}\right\}\xi_\al.\nn
\eea
}\hspace{-1.5mm}
By comparison with Eq. (\ref{LVAI-H}), there should not have any scalar potential coupling to the
``magnetic" part of $H$-coefficient, however, due to the tensor nature, the non-zero spatial metric $h_{ij}$ induce gravito-electric couplings to $\vec{\tilde{H}}$,
such as the terms proportional to $\phi_g\vec{\tilde{H}}\cdot\vec{\sigma}$ and $(\phi_g\vec{\tilde{H}}\cdot\vec{l})(\vec{l}\cdot\vec{\si})$.

A striking difference from the fermion-photon interaction is the presence of $a$-coupling terms
in (\ref{GI-ab1}) and (\ref{GFab2}).
Comparing (\ref{GFab2}) with (\ref{LIGI0}), we see the $a^\mu$ coefficient couples to $\phi_g$ and $\vec{A}_g$ in exactly the same way as the 4-momentum $p^\mu$.
This is not surprising as in the momentum space
\bea&&
-\hf\,h_{ba}\psb\,a^b\ga^a\ps+\frac{i}{4}h^b_{~a}\psb\ga^a\lrprt b\ps\Rightarrow\nn&&
~~~~~~~-\hf\,h_{ba}\bar{u}_\be(p')[a^b+\frac{p'^b+p^b}{2}]\ga^au_\al(p),\nonumber
\eea
and is also the same reason that $a^\mu$ can be shifted away by a phase redefinition of the fermion field,
thus does not have any observable consequence for a single fermion coupled with photon field in flat space.
However, the above reasoning does not apply for a fermion coupled with gravity \cite{Prospects2008}.
This can be verified by inspecting Eq. (\ref{LIGI0}),
where the simple replacement $p^\mu\rightarrow\,(p+a)^\mu$ cannot lead to the $a$-coupling terms in (\ref{GI-ab1}).

Note that we also need to consider the implicit correction to fermion-gravity interaction energy
induced by LV dispersion relation $p^0=\omega_0(\vec{p},~m)+\delta\omega(\vec{p},~m, X)$.
These correction comes from the substitution of $\vec{p}\cdot\vec{q}$ in the superficially LI term
$\frac{i}{4}\int\,d^3x\,h^b_{~a}\psb\ga^a\lrprt b\ps$, just as what we did in Eq. (\ref{deltaWX}).
However, in the gravitational case, additional contribution comes from $p^0$ term in (\ref{LIGI0}),
and thus is proportional to $\delta\omega_p$.
Inspection of $\delta\omega_p-\delta\omega_{p'}$ for various LV coefficients in the Appendix \ref{LVeigenSpinors},
we see these terms are at least of $\mathcal{O}(v)$, so in making a substitution of
$\vec{q}\cdot\vec{p}=-\frac{\vec{q}^2}{2}+m[\delta\omega_p-\delta\omega_{p'}]$ and $p^0=\omega_0+\delta\omega_p$,
the following correction
\bea\label{GI-deltaO}&&
\phi_g\left[(1+\frac{\vec{p}^2+i\vec{q}\times\vec{p}\cdot\vec{\si}}{4m^2})\delta\omega_p
+\frac{5}{4}(\delta\omega_p-\delta\omega_{p'})\right]\nn&&~~~~
-\left[\frac{\vec{A}_g\cdot\vec{l}+i\vec{q}\times\vec{A}_g\cdot\vec{\si}}{4m}\right]\delta\omega_p
\eea
has to be added for each type of LV coefficient.
For completeness, the dispersion relation for $a$-coefficient is $k^0=\sqrt{(\vec{k}+\vec{a})^2+m^2}-a^0$ for a positive energy fermion,
and hence $\delta\omega_p\simeq{\vec{k}\cdot\vec{a}}/{\omega_0}-a^0$.
For $a,~b,~H$-coefficients, the corrections due to LV dispersion relations are listed below,
{\small
\bea\label{DSC-ab}&&
\hspace{-5mm}E_\mathrm{GI}^\mathrm{ab-3}=\int\,d^3x\,e^{-i\vec{q}\cdot\vec{x}}\xi_\be^\dagger\left\{
\phi_{g}\left(\frac{(4\vec{p} - 5\vec{q}) \cdot (\vec{a}+b^{0}\vec{\si})}{4m}\right)+(a^0
\right.\nn&&
\hspace{-3mm}\left.
+\vec{b}\cdot\vec{\sigma})\left(\frac{\vec{A}_{g} \cdot \vec{l}+\vec{\Omega} \cdot \vec{\sigma}}{4 m}
-\left[\phi_g+\frac{\vec{g} \times \vec{p} \cdot \vec{\sigma}}{4 m^{2}} \right]\right)+\phi_g
\left[\frac{\vec{p}^{2}}{4m^{2}}\cdot
\right.\right.\nn&&
\hspace{-3mm}\left.\left.(\vec{b}\cdot\vec{\si}-a^0)+
\frac{5(\vec{p'} \cdot \vec{b})(\vec{p'} \cdot \vec{\sigma}) - 9(\vec{p} \cdot \vec{b})(\vec{p} \cdot \vec{\sigma})}{8m^2}
\right]\right\}\xi_\al,\\\label{DSC-H}&&
\hspace{-5mm}E_\mathrm{GI}^\mathrm{H-3}=
\int\,d^3x\,e^{-i\vec{q}\cdot\vec{x}}\xi_\be^\dagger\left\{\phi_{g}\left[\frac{\vec{H} \times (4\vec{p} - 5\vec{q}) \cdot \vec{\sigma}}{4m}  + \left(1+\frac{\vec{p}^{2}}{4 m^{2}}\right)\right.\right.\nn&&
\hspace{-3mm}\left.\left.
\vec{\tilde{H}} \cdot \vec{\sigma}+ \frac{5\vec{\tilde{H}} \cdot \vec{p'} \vec{\sigma} \cdot \vec{p'} - 9\vec{\tilde{H}} \cdot \vec{p}\, \vec{\sigma} \cdot \vec{p}}{8m^2}\right]
-\left[\frac{\vec{A}_{g} \cdot \vec{l}+\vec{\Omega} \cdot \vec{\sigma}}{4 m}\right.\right.\nn&&
\hspace{-3mm}\left.\left.
-\frac{\vec{g} \times \vec{p} \cdot \vec{\sigma}}{4 m^{2}}  \right] \vec{\tilde{H}} \cdot \vec{\sigma}
\right\}\xi_\al.
\eea
The total NR fermion-gravity interaction energy from $a~,b,~H$ contributions is the summation of Eqs.
(\ref{GI-ab1}-\ref{GFH2}) and (\ref{DSC-ab}-\ref{DSC-H}).
Though it is easy to see that several terms in the above equations can be combined together or even canceled,
such as the terms proportional to $\phi_g\vec{b}\cdot\vec{\si}$ and $a^0\frac{\vec{A}_{g} \cdot \vec{l}+\vec{\Omega} \cdot \vec{\sigma}}{4 m}$ in (\ref{GFab2})
and (\ref{DSC-ab}), or the terms proportional to $\vec{g}\times\vec{a}\cdot\vec{\si}$ in (\ref{GI-ab1}) and (\ref{GFab2}),
we keep them separately for the clarity of their origin.

Inspecting Eq. (\ref{GI-ab1}-\ref{GFH2}) and (\ref{DSC-ab}-\ref{DSC-H}) reveals that there are abundant interaction structures
for the LV spin-gravity coupling, especially for the $b,~H$ coefficients.
For example, the $-\frac{(\vec{\tilde{H}}+\vec{b})\cdot\vec{\Omega}}{4m}$ term is in analogy with the LV magnetic field coupling term $\frac{g\vec{b}\cdot\vec{B}}{2m^2}$ in (\ref{LVAI-b}), but with gravito-magnetic field $\vec{\Omega}$ replacing the magnetic field $\vec{B}$.
Similarly, the $\frac{(\vec{b}-\vec{\tilde{H}})\cdot \vec{\sigma} (\vec{\Omega} \cdot \vec{\sigma})}{4 m}$ and the spin-orbit coupling terms
such as those proportional to $\frac{\vec{g} \times \vec{p} \cdot \vec{\sigma}}{4 m^{2}}$ alter the geodetic and frame-dragging precession frequencies
of microscopic particles. Since there is no reason for the LV coefficients to be universal for particles with different flavor, the weak EP must be violated due to the non-universal LV gravitational couplings. These effects are in principle testable, such as in the high precision Gravity Probe B-like experiment \cite{GPB1}\cite{GPB2}.

Aside from the $B$-type LV couplings, the $E$-type LV couplings also show some similarity between the fermion-photon and fermion-gravity couplings,
such as $-\frac{\vec{E}\cdot\vec{H}}{4m^2}$ and $-\frac{\vec{g}\cdot\vec{H}}{4m}$, or $\frac{ig\vec{E}\times\vec{H}}{2m^2}\cdot\vec{\si}$ and $\frac{-2i\vec{g}\times\vec{H}}{m}\cdot\vec{\si}$.
The similarities for the LV couplings between the $b,~H$ coefficients can be trace back to the operator level by the identity $\psb\ga_5\vec{\ga}\ps=-\psb\ga_0\vec{\Si}\ps$, while $\ga^0$ is effectively equal to $\hat{1}$ for positive energy particles.
For example, this fact can be validated by the similar form of couplings between $\vec{b} \cdot \vec{\sigma}$ and $\vec{\tilde{H}} \cdot \vec{\sigma}$
with $[\vec{\Omega}-{\vec{g}\times\vec{p}}/{m}]\cdot\vec{\sigma}$ in (\ref{DSC-ab}) and (\ref{DSC-H}).

Next, we discuss the fermion-gravity interaction energies due to the $c,~d$ coefficients.
The contributions due to eigen-spinor corrections for $c,~d$ coefficients are
\bea
\label{GI-c1}&&
\hspace{-3mm}E_\mathrm{GI}^{c-1}=\frac{1}{4}\int\,d^3x\,e^{-i\vec{q}\cdot\vec{x}}\xi^\dagger_\be
\left\{[A^k_gl^0-2\phi_gl^k][c_{ij}\frac{\delta_{ik}l^j+i\epsilon_{ikl}q^j\si^l}{2m}]\right.\nn&&
\left.
+[\vec{l}\cdot\vec{A}_g-2l^0\phi_g][\frac{c_{(ij)}p'^ip^j+ic_{ij}q^{[j}p^{k]}
\epsilon_{ikl}\si^l}{2m^2}]\right\}\xi_\al,\\
\label{GI-d1}&&
\hspace{-3mm}E_\mathrm{GI}^{d-1}=\frac{1}{4}\int\,d^3x\,e^{-i\vec{q}\cdot\vec{x}}\xi^\dagger_\be
\left\{[2l^0\phi_g-\vec{l}\cdot\vec{A}_g]\frac{d_{0j}p'^{(i}p^{j)}\si^i}{2m^2}\right.\nn&&
\left.
+[2\phi_gl^m-A_g^ml^0]\left[\frac{d_{0j}l^j}{2m}\si^m+\frac{id_{ji}\epsilon_{jkm}(p^kp^i-p'^kp'^i)}{2m^2}
\right.\right.\nn&&
\left.\left.
+\frac{d_{mi}\si^k(p^kp^i+p'^kp'^i)-d_{ji}\si^j(p^mp^i+p'^mp'^i)}{2m^2}\right]\right\}\xi_\al
\eea
Compared the terms in (\ref{GI-c1}) and (\ref{GI-d1}) taking the form of $(A_g^k\,{X_k}-2\phi_g\,Y)l^0$, where $X_k, Y$ are LV operators such as $\frac{d_{0j}l^j}{2m}\si^k, -\frac{c_{ij}g^{[j}p^{k]}\epsilon_{ikl}\si^l}{2m}$, with the terms in (\ref{AIES-c2}) and (\ref{AIES-d2}), we see they also look quite similar, as mentioned in the general discussion of Eq. (\ref{GFSpinLV}).

The apparently LV vertex contributions due to $c,~d$ coefficients are
\bea\label{GIcdV}&&
E_\mathrm{GI}^\mathrm{cd-V}=
-\frac{i}{4}\langle p',\be|\int\,d^3x\,\left\{h^b_{~a}\left[\bar{\psi}(c_e^{~a}+d_e^{~a}\ga_5)\ga^e\lrprt b\ps\right]\right.
\nn&&\left.
~~~~~~~~+h^{\nu a}\left[\bar{\psi}(c_{b\nu}+d_{b\nu}\ga_5)\ga^b\lrprt a\ps\right]\right.
\nn&&\left.
~~~~~~~~+h^{\rho}_{~b}\left[\bar{\psi}(c_\rho^{~a}+d_\rho^{~a}\ga_5)\ga^b\lrprt a\ps\right]\right\}|p,\al\rangle.
\eea
Note the terms in the first line are in fact equal to the terms in the second line of (\ref{GIcdV}).
The terms in the 3rd line of Eq. (\ref{hIntGMS}),
$-\frac{1}{4}\epsilon^{bcmn}
\langle p',\be|\int\,d^3x\,h_{am,n}\psb[c_b^{~a}\ga_5+d_b^{~a}]\ga_c\ps|p,\al\rangle$
comes from the spin-connection interaction, and thus only contains GEM field strength $\prt_\rho h_{\mn}$,
and is naively expected to be much smaller than the terms coupled directly with the metric perturbation $h_{\mn}$.
\begin{widetext}
While the contributions due to LV dispersion relation corrections for the $c,~d$ coefficients are
\bea\label{DSC-c}&&
\hspace{-7mm}E_\mathrm{GI}^\mathrm{c-2}=
\int\,d^3x\,e^{-i\vec{q}\cdot\vec{x}}\xi_{\beta}^{\dagger}\left\{
c_{00} \left[\frac{\vec{A}_{g} \cdot \vec{l}+\vec{\Omega}\cdot \vec{\sigma}}{4}-\frac{\vec{g} \times \vec{p} \cdot \vec{\sigma}}{4 m}\right]+\phi_{g}
\left[c_{(0 j)}\frac{5 q^{j} - 4 p^{j} }{2}+ c_{(i j)} \frac{5 p'^{i} p'^{j} - 9 p^{i} p^{j}}{4m} -c_{00}(m+\frac{3\vec{p}^{2}}{4 m})\right]\right\}\xi_{\alpha},
\\\label{DSC-d}&&
\hspace{-7mm}E_\mathrm{GI}^\mathrm{d-2}=
\int\,d^3x\,e^{-i\vec{q}\cdot\vec{x}}\xi_{\beta}^{\dagger}\left\{\phi_{g}\left[\frac{9(\vec{p} \cdot \vec{d})(\vec{p} \cdot \vec{\sigma}) - 5(\vec{p'} \cdot \vec{d})(\vec{p'} \cdot \vec{\sigma})}{2m}
+ \frac{\tilde{d}_{ji} (4p^{i} - 5q^{i})\sigma^{j}}{4} + \frac{ 5d_{j0} p'^{j} \vec{\sigma} \cdot \vec{p'} - 9d_{j0} p^{j} \vec{\sigma} \cdot \vec{p}}{8m}
\right.\right.\nn&&\left.\left.~~
+ d_{j0}\sigma^{j}(m+\frac{\vec{p}^{2}}{4 m})\right]
+\frac{\vec{g} \times \vec{p} \cdot \vec{\sigma}}{4 m}d_{j 0}\sigma^{j}
-\frac{\vec{A}_{g} \cdot \vec{l}+\vec{\Omega} \cdot \vec{\sigma}}{4 } d_{j 0} \sigma^{j}\right\}\xi_{\alpha}.
\eea
In comparison with Eq. (\ref{AICur-c}), there is also a parallel term $c_{00}(\vec{A}_{g} \cdot \vec{l}+\vec{\Omega}\cdot \vec{\sigma})/{4}$,
which rescales the gravito-magnetic moment just as the corresponding term rescales the magnetic moment. This and the other similar LV corrections
spoil the theorem of null anomalous gravito-magnetic moment due to the equivalence principle \cite{Okun1963},
which is not unexpected in the LV theory.

For compactness, we combine these $c,~d$ couplings together and disregard the quadratic terms of $q^i$, as $q^iq^jh_{0\nu}\sim\prt_i\prt_jh_{0\nu}$
and $|\prt_i\prt_jh_{0\nu}|\ll|p^i\prt_jh_{0\nu}|$ in general. We also keep only terms up to $\mathcal{O}(m^{-1})$, and then the results are
{\small
\bea\label{GI-cT}&&
\hspace{-5mm}E_\mathrm{GI}^\mathrm{c}=\int\,d^3x\,e^{-i\vec{q}\cdot\vec{x}}\xi_{\beta}^{\dagger}
\left\{\left[
c_{00}\left(\frac{\vec{g} \times \vec{p} \cdot \vec{\sigma}}{2 m}+ 2m\phi_{g}\right) +c_{(0 i)} \phi_{g} \frac{7 q^{i}}{2} -l^{i} \phi_{g} c_{0 i} + \frac{1}{2}\epsilon_{i j k}c_{0 i} g^{k} \sigma^{j}  - c_{i j}\epsilon_{i k l}\left( \frac{3g^{(j} p^{k)}}{2} +
\frac{ p^{j}g^k }{2} \right)\frac{ \sigma^{l} }{ m} \right.\right.\nonumber\\&&
\left.\left.
- c_{(i j)}\phi_g \frac{2 p^{i} q^{ j}}{ m} -\epsilon_{i j k} c_{i j} g^{k} \frac{\vec{\sigma} \cdot \vec{p}}{4 m}\right]
+\left[m A_{g}^{i} \left(\frac{c_{0 i}}{2} + c_{(0 i)}\right) -c_{00}\left(\vec{A_g} \cdot \vec{p}+ \frac{\vec{\Omega} \cdot \vec{\sigma}}{4}\right)
+ c_{i j} \frac{A_{g}^{i} l^{j} +i \epsilon_{i k l} q^{j} A_{g}^{k} \sigma^{l}}{4} +\frac{l^{i} c_{i j}}{4}  A_{g}^{j}
\right.\right.\nonumber\\&&
\left.\left.
+ \left[l^{i}-\epsilon_{ikl}\sigma_l\partial_k \right] A_{g}^{j} \frac{c_{i j} }{2}
-\epsilon_{i j k}\frac{c_{i l} }{4} \partial_{k} A_{g}^{l} \sigma^{j}\right]
\right\}\xi_{\alpha},
\\\label{GI-dT}&&
\hspace{-5mm}E_\mathrm{GI}^\mathrm{d}=\int\,d^3x\,e^{-i\vec{q}\cdot\vec{x}}\xi_{\beta}^{\dagger}
\left\{\phi_{g}\left[d_{i j}\left(4 p^{j}+ \frac{q^{j}}{4}\right)\sigma^{i} - d_{00} \frac{\vec{\sigma} \cdot \left(8\vec{p}+11 \vec{q}\right)}{4}
-\frac{7 d_{j 0} q^{(i} p^{j)} \sigma^{i} }{4 m}\right] + \hf \epsilon_{ijk} d_{ij} g^{k}
+ \left(d_{i 0}+2d_{( 0 i )}\right)\frac{(\vec{g}\times\vec{p})^{i}}{2 m}
\right.\nn&&
\left.+ \frac{id_{j0}}{4 m}  [\vec{\sigma}\times(\vec{g} \times \vec{p})]^j
+m d_{00}\frac{\vec{A}_{g}\cdot\vec{\si}}{2}  + \frac{d_{i 0}}{4}\left(\sigma^{i}\vec{A}_{g}- A_{g}^{i}\vec{\sigma} \right)\cdot \vec{l} - d_{0 i} A_{g}^{i} \frac{\vec{\sigma} \cdot \vec{l}}{2} -\frac{i\,d_{j 0}}{4}(\vec{\Omega} \times \vec{\sigma})^j  - m A_{g}^{j} d_{i j} \sigma^{i}
\right\}\xi_{\alpha}.
\eea
}
\end{widetext}

Inspection of (\ref{DSC-c}-\ref{GI-dT}) shows that several LV spin-orbit coupling terms, such as
$d_{j0}\si^j{\vec{g}\times\vec{p}\cdot\vec{\sigma}}/{4m}$, $c_{00}{\vec{g}\times\vec{p}\cdot\vec{\sigma}}/{2m}$ and $c_{ij}(\vec{g}\times\vec{\si})^ip^j/2m$,
are of the similar kind of structure as we found in FE interactions, like $c_{ij}\frac{(\vec{E}\times\vec{\si})^ip^j-(\vec{p}\times\vec{\si})^iE^j}{4m^2}$.
Other more complicated structures of spin-orbit couplings, such as $\frac{c_{0i}}{2}(\vec{g}\times\vec{\si})^i$, $-\frac{i\,d_{j 0}}{4}(\vec{\Omega} \times \vec{\sigma})^j$, $d_{(0i)}A_{g}^{i}{\vec{\sigma}\cdot\vec{l}}/{2},  \frac{id_{j0}}{4 m}  [\vec{\sigma}\times(\vec{g} \times \vec{p})]^j$, \etc.
can also be found in Eq. (\ref{GI-cT}) and (\ref{GI-dT}).
Also we notice that there are only two spin-independent fermion-gravity coupling for the $d$ coefficient, $(d_{i0}+2d_{(0i)})\frac{(\vec{g}\times\vec{p})^{i}}{2m}$ and $\hf \epsilon_{ijk} d_{ij} g^{k}$.
This is not surprising as in the Lagrangian level, $d$ term is of the $\ga_5\ga^a$ structure and is an essentially spin-dependent term
from the relativistic point of view.

In summary, due to similar Dirac structures in Lorentz violating fermion-gravity (FG) and fermion-electromagnetic (FE) couplings,
there are analog operators for the LV fermion couplings with these two external fields.
As a simple glance, we collect several sampling operators in FG and FE interactions in table \ref{AnaOper}.
Operators such as $a^0\frac{\vec{A}_{g}\cdot\vec{l}+\vec{\Omega}\cdot\vec{\si}}{4m}$, $\frac{(\vec{A}_g\times\vec{H})\cdot\vec{\si}}{2}$ exactly cancel, and thus in fact do not appear. The mismatch between FG and FE interactions may partly due to the tensor structure of gravity, and
partly due to the fact that the LV corrections from fermion dispersion relations $p^0=\omega^0+\delta\omega_p$ can contribute directly in the case of gravity,
in contrast to the case of photon coupling, where only $\delta\omega_p-\delta\omega_{p'}$ enters in the $q\cdot p$ subsititution.
Anyway, we think even the sample operators in table \ref{AnaOper} can convince the readers that the LV spin coupling structures are very abundant, which means the very rich gravitational phenomenologies arising from the LV spin-gravity couplings \cite{Kost-BeyondRieman} are waiting for us to explore.

\begin{table}[ht]
\begin{center}
\renewcommand{\arraystretch}{1.7}
\begin{tabular}{c|c|c|c|c|c|c|c|c}
\hline\hline
   FG    & $-\frac{(\vec{\tilde{H}}+\vec{b})\cdot\vec{\Omega}}{4m}$  & $\frac{(\vec{A}_g\cdot\vec{l})(\vec{b}\cdot\vec{\si})}{2m}$ &  $\phi_g\frac{\vec{p}\times\vec{H}\cdot\vec{\si}}{2m}$ &  $\frac{(\vec{A}_g\cdot\vec{\si})(\vec{p}\cdot\vec{\tilde{H}})}{m}$ \\\hline
   FE    &    $\frac{g\vec{b}\cdot\vec{B}}{2m^2}$  &  $\frac{g(\vec{A}\cdot\vec{l})(\vec{b}\cdot\vec{\si})}{2m^2}$ & $-\frac{gA^0\vec{p}\times\vec{H}\cdot\vec{\si}}{2m^2}$  &
    $-\frac{g(\vec{A}\cdot\vec{\si})(\vec{p}\cdot\vec{\tilde{H}})}{m^2}$ \\
 \hline \hline
   FG     &  $-\frac{c_{00}(2\vec{A}_{g}\cdot\vec{l}+\vec{\Omega}\cdot \vec{\sigma})}{4}$   & $\frac{3c_{ij}g^{[j}p^{k]}\epsilon_{ikl}\si^l}{2m}$ &  $-\frac{\vec{g}\cdot\vec{H}}{4m}$ & $-\frac{(\vec{d}\cdot\vec{A}_g)(\vec{\sigma}\cdot\vec{l})}{2}$   \\\hline
   FE     &  $\frac{g\,c_{00}(\vec{A}\cdot\vec{l}+\vec{B}\cdot \vec{\sigma})}{2m}$   &  $\frac{g\,c_{ij}E^{[j}p^{k]}\epsilon_{ikl}\si^l}{2m^2}$ & $-\frac{g\vec{E}\cdot\vec{H}}{4m^2}$ &
   $\frac{g(\vec{d}\cdot\vec{A})(\vec{\sigma}\cdot\vec{l})}{m}$  \\
\hline\hline
\end{tabular}\caption{Examples of the analogous couplings between LV fermion-gravity and fermion-photon couplings. }\label{AnaOper}
\end{center}
\end{table}

\section{Phenomenology in test of EP}\label{ConstrEP}
The LV spin-gravity couplings have already been thoroughly explored in the uniform limit $\phi_g=\vec{g}\cdot\vec{z}$ \cite{Kost-BeyondRieman},
which is a very good approximation for most experiments on the Earth.
However, linear potential is an essentially flat metric, and is incapable to capture the warp effects of space, as only $g_{00}$ really matters in this case.
In comparison, the Lense-Thirring metric is an intrinsically curved one, and may be able to test LV spin-gravity couplings where the other metric components take effect, such as the frame-dragging (FD) effect of a single fermion due to the rotation of a massive object like neutron star.
For the pure gravity sector, we also note that the spin precession effects in the post Newtonian approximation up to $\mathcal{O}(3)$
have already been systematically studied \cite{SignalLVPN},
and the anomalous precession rates due to LV have also been utilized to constrain the $s_{\mn}$ coefficients \cite{LimitS-GPB13}.
However, these are for macroscopic spinning gyroscopes, not for the intrinsic spin of microscopic fermions.

The Lorentz invariant NR fermion-gravity Hamiltonian has been fully studied in the literature \cite{Fischbach}\cite{WTHehl1990}\cite{Obukhov2009}, and it is interesting to note that the LI operators in $E_\mathrm{GI}^\mathrm{LI}$, Eq. (\ref{LIGI0}), coincide with those in the NR fermion Hamiltonian obtained in \cite{Fischbach} except the higher order term $\phi_g\vec{p}^2/2m$, which differ by an $\mathcal{O}(1)$ numerical factor.
The vanishing of terms $\nabla^2\phi_g,~\frac{i\vec{g}\cdot\vec{p}}{m}$ is due to our on-shell and source free assumptions.
This is not surprising, as the LI terms in the one-fermion matrix element $-\hf\langle p',\be|\int\,d^3x\,h_{\mn}T^{\mn}|p,\al\rangle$ under the assumption
of zero energy transfer $q^0=0$ is just the potential energy in tree level approximation.
For the LV counter terms, we may also expect them to be the corresponding LV operators in the NR Hamiltonian obtained by Foldy-Wouthuysen transformation \cite{FWT1950}\cite{Tassonspin}\cite{ZX2018},
except each pair of operators obtained from different approaches may differ by an $\mathcal{O}(1)$ numerical factor.
As most LV coefficients in the minimal SME have been tightly constrained to be very vanishingly small \cite{dataTable},
what we really cared about is essentially the order of magnitude,
the $\mathcal{O}(1)$ numerical factors may be irrelevant for practical purposes.
Thus we can collect all spin-dependent operators up to $\mathcal{O}(m^{-1})$ (except the $a^0$ term)
\bea&&
\hspace{-3mm}\delta\hat{H}_{g\si}=\left[\frac{3\vec{g}\times\vec{a}}{4m}-\frac{a^0\vec{g}\times\vec{p}}{4m^2}\right]
\cdot\vec{\si}+\phi_g\left[\frac{3b^0\vec{p}}{2m}-2\vec{b}\right]\cdot\vec{\si}\nn&&
+\frac{\vec{\si}\cdot\vec{p}}{m}\vec{A}_g\cdot\vec{b}
+\phi_g\left[\frac{\vec{p}\times\vec{H}}{2m}+3\vec{\tilde{H}}\right]\cdot\vec{\si}
+\frac{\vec{A}_g\cdot(\vec{\si}\vec{p}-\vec{p}\vec{\si})\cdot\vec{\tilde{H}}}{m}\nn&&
+c_{00}\left[\frac{\vec{g}\times\vec{p}}{2m}-\frac{\vec{\Omega}}{4}\right]\cdot\vec{\si}
+d_{00}\left[\frac{m\vec{A}_g}{2}-2\phi_g\vec{p}\right]\cdot\vec{\si}
\eea
together, and for simplicity we also ignore the terms coupled with $c_{\mn},~d_{\mn}$ coefficient, except the $c_{00}$ and $d_{00}$.
We boldly assume the NR Hamiltonian is
\bea\label{SpinNR}&&
\hat{H}_\mathrm{NR}=\frac{\vec{p}^2}{m}+m\phi_g+\frac{3}{2m}\left[\phi_g\vec{p}^2
-i\vec{g}\cdot\vec{p}+\vec{g}\times\vec{p}\cdot\frac{\si}{2}\right]-\frac{\vec{\Omega}\cdot\vec{\si}}{4}\nn&&~~~~
+\delta\hat{H}_{g\si},
\eea
where the first line are LI contributions.
Note we have ignored all the spin-independent LV operators, as they do not directly affect spin dynamics.
The spin time evolution is governed by the Heisenberg equation
\bea\label{Spinpre1}
\frac{d\vec{S}}{dt}=\frac{1}{i\hbar}[\vec{S},~\hat{H}_\mathrm{NR}]=
(\vec{\omega}_{_\mathrm{LI}}+\delta\vec{\omega}_{_\mathrm{LV}})\times\vec{S},
\eea
where $\vec{\omega}_{_\mathrm{LI}}\equiv\vec{\omega}_{_\mathrm{geo}}+\vec{\omega}_{_\mathrm{FD}}$,
and $\vec{\omega}_{_\mathrm{geo}}=\frac{3}{2m}\vec{g}\times\vec{p}$ and $\vec{\omega}_{_\mathrm{FD}}=-\frac{\vec{\Omega}}{2}$ describe the geodetic precession and
FD precession angular vectors predicted in GR, respectively.
It is interesting that we obtain $\vec{\omega}_{_\mathrm{LI}}$ from fermion-gravity couplings
[and maybe by accident the LI spin interaction terms do have the correct numerical factors,
while this is not so for spin-independent terms.
For \eg, the numerical factor in front of $\phi_g\frac{\vec{p}^2}{m}$ in Eq. (\ref{LIGI0}) does not
coincide with the one obtained by the FW transformation],
which only relies on the minimal fermion-gravity couplings within the tetrad formalism.
This can be viewed as an evidence that the WEP is valid even in the quantum regime \cite{SemiClaGF-Silenko}\cite{Obukhov2009}\cite{COW-n}.
Given that WEP has been tested to high precision \cite{COW-n}\cite{AITEPe12}\cite{SpinEPe7}, we may reasonably believe that the GR predicted spin precession of microscopic particles
can also be tested to the same precision in the future as that of the macroscopic gyroscope in the famous Gravity Prob B (GPB) project \cite{GPB1},
in addition to technique difficulties caused by the extremely weak fermion-gravity couplings.
The GPB gives a geodetic drift rate of $R_\mathrm{NS,o}=6601.8\pm18.3$ mas/yr and a frame-dragging drift rate of $R_\mathrm{WE,o}=37.2\pm7.2$ mas/yr, while the corresponding drift rates predicted by GR are of $R_\mathrm{geo}=6606.1$ mas/yr and $R_\mathrm{FD}=39.2$ mas/yr, respectively,
so the measured drift rate deviations are $|\Delta R_\mathrm{NS}|<22.6$ mas/yr and $|\Delta R_\mathrm{WE}|<9.2$ mas/yr \cite{LimitS-GPB13}.
The LV induced anomalous precession is
\bea&&
\hspace{-3mm}
\delta\vec{\omega}_{_\mathrm{LV}}=\frac{\vec{g}}{2m}\times\left[3\vec{a}-\frac{a^0\vec{p}}{m}\right]
+\phi_g\left[\frac{3b^0\vec{p}}{m}+(6\vec{\tilde{H}}-4\vec{b})\right]
\nn&&~~~~
+\frac{2\vec{A}_g\cdot\vec{b}}{m}\vec{p}+\phi_g\frac{\vec{p}\times\vec{H}}{m}
+\frac{2}{m}\left[(\vec{\tilde{H}}\cdot\vec{p})\,\vec{A}_g-(\vec{A}_g\cdot\vec{p})\,\vec{\tilde{H}}\right]
\nn&&~~~~
c_{00}\left[\frac{\vec{g}\times\vec{p}}{m}-\frac{\vec{\Omega}}{2}\right]
+d_{00}\left[m\vec{A}_g-4\phi_g\vec{p}\right].
\eea
If we attribute all the drift rate deviations to the LV caused anomalous precession and assume that
the same precision can be achieved for fermion spin precession measurement,
we may obtain some very rough bounds on
\bea\label{Hbbounds}&&
\hspace{-3mm}|3\vec{\tilde{H}}-2\vec{b}|\leq\frac{3\Delta R_\mathrm{NS}}{4R_\mathrm{geo}}\frac{v}{r}
\simeq5.432\times10^{-13}\mathrm{eV}\\\label{acbounds}&&
\hspace{-3mm}|a^0|\leq3m\frac{\Delta R_\mathrm{NS}}{R_\mathrm{geo}}\simeq9.65\times10^6\mathrm{eV},\\\label{cbounds}&&
\hspace{-3mm}|c_{00}|\leq\mathrm{Min}\{\frac{3}{2}\frac{\Delta R_\mathrm{NS}}{R_\mathrm{geo}},\frac{\Delta R_\mathrm{WE}}{R_\mathrm{FD}}\}=5.14\times10^{-3},
\eea
where we set $r=7018.0$km as the GPB polar orbit parameter (orbit altitude $642$km) and assume each type of LV coefficient as the only non-zero one in our calculations.
The bounds are weak as they are obtained from the deviation of the essentially weak GR effects.
Also note we intentionally choose the above LV coefficients as our naive estimates,
cause the other LV operators such as $\vec{g}\times\vec{a},~b^0\vec{p},~\vec{p}\times\vec{H}$ may be even weaker as they may average out in an evolution, not mention the data acquisition period is almost 1 year, from August 2004 to August 2005. In other words, if we had transformed to the Sun-centered frame, our estimates could be even weaker.
The LAGEOS, LAGEOS 2 and LARES laser-ranged satellites can test the LT nodal shift to the accuracy $0.2\%$ \cite{LLLC2016}, and this in principle may put at least 2 order of magnitude more tighter bounds to the LV coefficients, though it is more unlikely as the test is not even for a gyroscope on an orbit.
Another point is that our estimates are based on the assumption that fermion precession can be tested to the same accuracy as the macroscopic gyroscope. This means our bounds above are best to be viewed as perspectives.

If we consider the acceleration
\bea\label{acceleration}&&
\hspace{-5mm}\vec{a}\equiv\frac{d\vec{p}}{m\,dt}=\frac{1}{i\,m}[\vec{p},~\hat{H}_\mathrm{NR}]
\simeq-\nabla\phi_g(1+\frac{3\vec{p}^2}{2m^2})\nn&&
-\nabla\phi_g\left[\frac{(3\vec{\tilde{H}}-2\vec{b})}{m}-\frac{2d_{00}}{m}\vec{p}\right]\cdot\vec{\si},
\eea
where we have ignored all the LV corrections with higher order than $m^{-1}$ and the LV corrections coupled with
gravito-magnetic vector potential $\vec{A}_g$ or derivatives of $\vec{g}$, since we expect these terms to be
much tinier compared with the remained ones, and we note that the anomalous acceleration is purely due to the LV spin-gravity couplings.
We can then get bounds
\bea\label{Haccbound}&&
|3\vec{\tilde{H}}-2\vec{b}|\leq1.8\times10^{-7}m_{87}\simeq1.46\times10^{-5}\mathrm{GeV},\\\label{baccbound}&&
|d_{00}|\leq9\times10^{-8}\sqrt{\frac{m_{87}}{3k_B T}}\simeq4.51\times10^{-6},
\eea
from the test of weak EP with neutral atoms with the precision of
$\eta=(0.2\pm1.6)\times10^{-7}$ \cite{SpinEPe7}\cite{EPDuan}.
We choose the temperature as $T=1.4\mu$K \cite{SpinEPe7}, mass $m_{87}$ as the $87$ atomic mass unit,
as the particle involved are $^{87}\mathrm{Sr},^{88}\mathrm{Sr}$ and $^{87}\mathrm{Rb}$, roughly the same mass range, and $|\eta|\simeq1.8\times10^{-7}$, the most conserved one.
Since the time scale for two experiments are much smaller than a day, there is no need to take into account
of the sidereal variations for a rough estimate, and these weak bounds are more reliable.

\section{Summary}\label{Summ}
In this paper, we calculate the one-fermion matrix elements of fermion-electromagnetic (FE) and fermion-gravity (FG) interactions for on shell fermions.
Due to the partial structure similarities between FE and FG interactions, many LV fermion-gravity operators bare the resemblance to LV fermion-photon operators.
We have show the resemblance with several sampling operators in Table \ref{AnaOper}.
This resemblance can be viewed as a natural manifestation of the well-known gravito-electromagnetism generalized to the LV fermion couplings.

By collecting the spin-dependent LV operators in the matrix elements as leading order LV perturbation
and combined with the non-relativistic LI gravitational interaction in the Lense-Thirring (LT) metric, we obtain a hybrid Hamiltonian,
from which we obtain a spin precession equation (\ref{Spinpre1}) and a linear acceleration equation (\ref{acceleration}).
From the anomalous spin precession rate as the correction to the geodetic precession and LT frame-dragging precession predicted in
general relativity, we can get some weak bounds on gravitationally coupled LV fermion coefficients,
Eq. (\ref{Hbbounds}-\ref{cbounds}).
Though these constraints relies on an unrealistic assumption of the measurement capability, which says the
fermion-gravity coupling can be measured to the same precession as in Gravity Probe B project,
these bounds are interacting since they reveal another aspect of WEP test \cite{SemiClaGF-Silenko}\cite{Obukhov2009}, namely,
the spin precession of a microscopic fermion may be different from the macroscopic gyroscope if the LV
spin-gravity couplings are allowed.
From the WEP test with atoms of nonzero spin, we can also get some relatively stronger and more reliable bounds (\ref{Haccbound}-\ref{baccbound}) on
the LV fermion-gravity couplings.
These bounds do not require to take account of the sidereal effect induced by the Earth motion,
as the relevant time scale is much shorter than a sidereal day, however, the analysis of sidereal effect may necessarily puts the bound more stringent.
Moreover, the future high accuracy experiments with polarized neutral atoms may be able to give more tighter bounds on these LV spin-gravity couplings
\cite{GRAI2008}\cite{ELAGR2019}.

\section{Ackowledgement}
The author Z. Xiao appreciates the valuable discussions with Alan Kosteleck\'y and Tianbo Liu.
This work is partially supported by National Science Foundation of China under grant No. 11120101004,
No. 11974108, and the Fundamental Research Funds for the Central Universities under No. 2020MS042.

\appendix
\section{The Gravito-Electromagnetic Equations}\label{GraviEM}
The gravito-electromagnetism can be viewed as an analogy to electrodynamics when gravity is sufficiently weak for slow moving gravitational sources. For weak gravity, we can linearize the Einstein equation
$G_{\mn}=\kappa T_{\mn}$ by regarding the metric as a small deviation from Minkowski background
\bea\label{Metricdev}
g_{\mn}=\eta_{\mn}+h_{\mn},\quad |h_{\mn}|\ll1.
\eea
The field equation can be further simplified in the harmonic gauge $\prt_\mu\bar{h}^\mu_{~\nu}=0$ with trace reversed rank-2 tensor $\bar{h}_{\mn}=h_{\mn}-\frac{\eta_{\mn}}{2}h$ (where $h=\eta^{\mn}h_{\mn}$),
\bea\label{WaveEeqn}
\prt_\al\prt^\al\bar{h}_{\mn}=-2\kappa T_{\mn}, \quad \kappa\equiv\frac{8\pi G}{c^4}.
\eea
The a class of retarded solutions can be found as (\ref{WaveEeqn}) is simply a wave equation.
For $T^{00}\sim\rho_mc^2,~T^{0i}\sim\rho_mcu^i,~T^{ij}\sim\rho_mu^iu^j$, we get up to $\mathcal{O}(c^{-4})$,
$\bar{h}_{00}\equiv-\frac{4\phi_g}{c^2}$, $\bar{h}_{0i}=\frac{4A_g^i}{sc^{3-n}}$ and $\bar{h}_{ij}\sim\mathcal{O}(c^{-4})$, where
\bea\label{Grav4Vec}&&
\phi_g(x)=-G\int d^3y\frac{\rho_m(t-\frac{|\vec{x}-\vec{y}|}{c}; \vec{y})}{|\vec{x}-\vec{y}|},\\&&
A_g^i(x)=-\frac{s\,G}{c^n}\int d^3y\frac{\rho_m(t-\frac{|\vec{x}-\vec{y}|}{c}; \vec{y})u^i}{|\vec{x}-\vec{y}|}
\eea
Then define $\vec{E}_g\equiv-\nabla\phi_g-\frac{r}{c}\prt_t\vec{A}_g$ and $\vec{B}_g\equiv\nabla\times\vec{A}_g$. The $n,~s,~r$ are a set of constants to be determined.
It is easy to verify that the homogeneous equations
\bea\label{HomogeGEM}
\nabla\cdot\vec{B}_g=0, \quad \nabla\times\vec{E}_g=-\frac{r}{c}\prt_t\vec{B}_g
\eea
are satisfied automatically. Since the harmonic gauge $\prt_\mu\bar{h}^\mu_{~\nu}=0$ reads
\bea&&
0=\prt_j\bar{h}_{j0}-\prt_0\bar{h}_{00}=
\frac{4}{c^2}\left[\frac{\prt_jA_g^j}{s\,c^{1-n}}+\frac{1}{c}\prt_t\phi_g\right]\nn&&
0=\prt_0\bar{h}^0_{~i}+\prt_j\bar{h}^j_{~i}=-\prt_0\bar{h}_{0i}=-\frac{4\prt_tA_g^i}{s\,c^{4-n}},
\eea
the vector potential must be time independent, $\dot{A}_g^i=0$, and substituting
$\nabla\cdot\vec{A}_g=-\frac{s}{c^n}\prt_t\phi_g$ into the inhomogeneous equations gives
\bea\label{GravGauss}&&
\hspace{-5mm}\nabla\cdot\vec{E}_g=-\nabla^2\phi_g-\frac{r}{c}\prt_t(\nabla\cdot\vec{A}_g)=
-[\nabla^2-\frac{r\,s}{c^{n+1}}\prt_t^2]\phi_g\nn&&
\hspace{-3mm}\xlongequal[n=1]{r\,s=1}-\Box\phi_g=-4\pi G\rho_m,\\&&
\hspace{-5mm}\nabla\times\vec{B}_g=\nabla(\nabla\cdot\vec{A}_g)-\nabla^2\vec{A}_g
=-\frac{s}{c^n}\prt_t(\nabla\phi_g)-\nabla^2\vec{A}_g\nn&&
\hspace{-3mm}\xlongequal[n=1]{r\,s=1}\frac{s\prt_t\vec{E}_g}{c}-\Box\vec{A}_g
=s\left[\frac{\prt_t\vec{E}_g}{c}-\frac{4\pi G}{c}\rho_m\vec{u}\right].
\eea
where $\Box\equiv[\nabla^2-\frac{1}{c^2}\prt_t^2]$ is the flat space d' Alembert operator,
and in order to make use of $\Box_x\int d^3y\frac{f(t-\frac{|\vec{x}-\vec{y}|}{c}; \vec{y})}{|\vec{x}-\vec{y}|}=-4\pi f(t, \vec{x})$,
we have to set $r\,s=n=1$.

The geodesic equation $\frac{du^\al}{d\tau}+\Ga^\al_{~\be\ga}u^\be u^\ga=0$
can be written as
\bea\label{GeodesE}&&
 \left.\begin{array}{c}
\frac{du^\al}{dt}=-\Ga^\al_{~\be\ga}\frac{dx^\be}{dt}\frac{dx^\ga}{dt}\frac{dt}{d\tau}\\
\frac{du^\al}{dt}=\frac{d}{dt}[\frac{dx^\al}{dt}\frac{dt}{d\tau}]
=\frac{d^2x^\al}{dt^2}\frac{dt}{d\tau}+\frac{dx^\al}{dt}\frac{d^2t}{d\tau^2}\frac{d\tau}{dt}\\
  \end{array}\right\}\Rightarrow\nn&&
  \frac{d^2x^\al}{dt^2}
  =\left[\frac{1}{c}\frac{dx^\al}{dt}\Ga^0_{~\be\ga}-\Ga^\al_{~\be\ga}\right]\frac{dx^\be}{dt}\frac{dx^\ga}{dt}
\eea
Note in the weak gravitational field limit,
\bea\label{ChristWL}&&
\hspace{-5mm}\Ga^0_{00}\simeq-\hf h_{00,0}=\frac{\prt_t\phi_g}{c^3},\quad
\Ga^0_{0j}\simeq-\hf h_{00,j}=\frac{\prt_j\phi_g}{c^2},\nn&&
\hspace{-5mm}\Ga^0_{jk}\simeq\hf(h_{jk,0}-h_{0j,k}-h_{0k,j})=-\delta_{jk}\frac{\prt_t\phi_g}{c^3}
-\frac{2}{sc^2}({A^g}_{j,k}+{A^g}_{k,j}),\nn&&
\hspace{-5mm}\Ga^i_{00}\simeq[h_{i0,0}-\hf h_{00,i}]=\frac{4}{sc^3}\prt_t{A_g}^i+\frac{\prt_i\phi_g}{c^2},\nn&&
\hspace{-5mm}\Ga^i_{0j}\simeq\hf[h_{i0,j}+h_{ij,0}-h_{0j,i}]=\frac{2}{sc^2}\left[{A^g}_{i,j}-{A^g}_{j,i}\right]
-\delta_{ij}\frac{\prt_t\phi_g}{c^3},\nn&&
\hspace{-5mm}\Ga^i_{jk}\simeq\hf(h_{ij,k}+h_{ik,j}-h_{jk,i})=\frac{1}{c^2}[\delta_{jk}\prt_i
-\delta_{ik}\prt_j-\delta_{ij}\prt_k]\phi_g,
\eea
where $h_{ij}=-\frac{2\phi_g}{c^2}\delta_{ij},~h_{0j}=\frac{4A_g^j}{sc^2},~h_{00}=-\frac{2\phi_g}{c^2}$.
Substituting the above equations into the geodesic equation (\ref{GeodesE}), we get
\bea\label{geoaccel}&&
\hspace{-5mm}a^i\equiv\frac{d^2x^i}{dt^2}=[\frac{v^i}{c}\Ga^0_{~00}-\Ga^i_{~00}]c^2
+2[\frac{v^i}{c}\Ga^0_{~0j}-\Ga^i_{~0j}]cv^j\nn&&
+[\frac{v^i}{c}\Ga^0_{~jk}-\Ga^i_{~jk}]v^jv^k
=\left[\frac{3v^i}{c^2}\prt_t+\frac{4v^i}{c^2}(\vec{v}\cdot\nabla)\right]\phi_g
\nn&&
-\left[(\frac{4}{sc}\prt_t{A_g}^i+\prt_i\phi_g)+\frac{4}{sc}v^j({A^g}_{i,j}-{A^g}_{j,i})\right]\nn&&
-\frac{\vec{v}^2}{c^2}\prt_i\phi_g+\mathcal{O}(c^{-3}),
\eea
In comparison, if we want to have an analogy to the Lorentz force law, we have to set
$\frac{4}{s}=r$, which is in contradict to the condition $r\,s=1$ in the Eq. (\ref{GravGauss}).
To compromise, we have to resort to stationary assumption, where $\phi_g,~\vec{A}_g$ are time-independent,
and then $\Box\rightarrow\nabla^2$. A convention is $r=1,~s=4$, and then the
\bea
a^i\equiv\,(1+\frac{\vec{v}^2}{c^2})E^i_g+(\vec{v}\times\vec{B}_g)^i.
\eea

\section{the linear LV Lagrangian density}\label{LinLVLd}
The original Dirac equation obtained from (\ref{matterAct}) is
\bea\label{FullDirac}&&
\left[ie^\mu_{~a}\left(\Ga^a{\vec{\nabla}_\mu}+\frac{i}{8}\omega_\mu^{~bc}\left[\si_{bc},~\Ga^a\right]\right)-M\right]\psi\nn&&
+\frac{i}{2}e^\mu_{~a}\left\{\prt_\mu\Ga^a+\omega_{\mu~c}^{~a}\Ga^c\right\}\psi=0.
\eea
Now consider the linearized LV fermion-gravity Lagrangian in metric perturbation $h_{\mn}$. The LV fermion-gravity Lagrangian is
\bea
\mathscr{L}_\mathrm{LV}=\frac{i}{2}e^\mu_{~a}\bar{\psi}\delta\Ga^a\,{\lrNrt\mu}\,\psi-\bar{\psi}\delta M\ps
=\mathscr{L}_{c,d}+\mathscr{L}_{a,b,H},
\eea
where $\delta\Ga^a\equiv\Ga^a-\ga^a$ and $\delta M\equiv M-m$.
For the $c,~d$-coefficients, the corresponding Lagrangian is
{\small
\bea&&
\mathscr{L}_{c,d}=-\frac{i}{2}e^\mu_{~a}\bar{\psi}\left(c_{\rho\nu}+d_{\rho\nu}\ga_5\right)\ga^be^{\nu a}e^\rho_{~b}\lrNrt\mu\ps\nn&&
\hspace{-3mm}=-\frac{i}{2}e^\mu_{~a}\bar{\psi}(c_{\rho\nu}+d_{\rho\nu}\ga_5)\ga^b\,e^{\nu a}e^\rho_{~b}\lrdrt\mu\ps+\frac{1}{8}e^\mu_{~a}\omega_\mu^{~cd}\nn&&
~\cdot\bar{\psi}\{(c_{\rho\nu}+d_{\rho\nu}\ga_5)\ga^b,\si_{cd}\}e^{\nu a}e^\rho_{~b}\ps\nn&&
\hspace{-3mm}\simeq
-\frac{i}{2}\bar{\psi}[c_b^{~a}+d_b^{~a}\ga_5]\ga^b\left[\lrdrt a-\hf h^\mu_{~a}\lrdrt\mu\right]\ps\nn&&
\hspace{-3mm}+
\frac{i}{4}\bar{\psi}\left[h^{\nu a}(c_{b\nu}+d_{b\nu}\ga_5)\ga^b
+h^{\rho}_{~b}(c_\rho^{~a}+d_\rho^{~a}\ga_5)\ga^b\right]\lrdrt a\ps\nn&&
\hspace{-3mm}+\frac{1}{4}\epsilon^{bcmn}h_{am,n}\psb[c_b^{~a}\ga_5+d_b^{~a}]\ga_c\ps,
\eea
}
\hspace{-1.5mm}while for the $a,~b,~H$-coefficients, the contributions to the Lagrangian are
\bea&&
\hspace{-5mm}\mathscr{L}_{a,b,H}=-\psb\delta M\ps\simeq-\psb\left[(a_a+b_a\ga_5)\ga^a+\hf H_{bc}\si^{bc}\right]\ps\nn&&~~
\hspace{-4mm}+\frac{h^\mu_{~a}}{2}\psb\left[(a_\mu+b_\mu\ga_5)\ga^a+\hf(H_{\mu b}\si^{ab}+H_{b\mu}\si^{ba})]\right]\ps.
\eea

\section{Field Redefinition Procedure}\label{FieldRedP}
The field redefinition matrix for the linearized Lagrangian
$\mathscr{L}_\psi=(1+\hf h)\left[\mathscr{L}_\mathrm{LI}+\mathscr{L}_\mathrm{LV}\right]$ is
$\hat{U}\equiv1-\hf\ga^0C^0\equiv1+\delta\hat{U}_0+\delta\hat{U}^h$, where
$\delta\hat{U}_0\equiv\hf(d_{b0}\ga_5-c_{b0})\ga^0\ga^b$ is the redefinition matrix
in flat space and $\delta\hat{U}^h=\delta\hat{U}_I^h+\delta\hat{U}_V^h$ is the additional contribution
due to gravity. The LI and LV pieces of $\delta\hat{U}^h$ are
\bea\label{RedUhLI}&&
\hspace{-5mm}\delta\hat{U}_I^h=-\frac{1}{4}(h+h_{0\mu}\ga^0\ga^\mu),\\\label{RedUhLV}&&
\hspace{-5mm}\delta\hat{U}_V^h=-\frac{1}{4}\left[h^{\nu 0}(c_{b\nu}-d_{b\nu}\ga_5)+h^\rho_{~b}(d_{\rho0}\ga_5-c_{\rho0})\right]\ga^0\ga^b\nn&&
~~~~
-\frac{1}{4}(h\,\ga^0\delta\Ga_\circ^0-h_{~\mu}^0\ga^0\delta\Ga_\circ^\mu),
\eea
respectively. The spinor redefinition is $\psi=\hat{U}\chi$ and the associated fermion bilinear
$\psb\hat{\mathcal{O}}\ps$ after spinor redefinition is $\bar{\chi}\ga^0\hat{U}^\dagger\ga^0\hat{\mathcal{O}}\hat{U}\chi$.
However, up to linear order approximation in $h_{\mn}$, there is an effective distinction between
\bit
\item
any operator constructed from the Lagrangian $\mathscr{L}_\psi$ linear in $h_{\mn}$.
There is no need to taking account of $\delta\hat{U}^h$,
as otherwise the resultant operator is of order $\mathcal{O}(h^2)$.
In other words, we only need to take $\hat{U}_0\equiv1+\delta\hat{U}_0$ as the redefinition matrix.
\item any ``flat space" operator such as $\frac{i}{2}\bar{\psi}\Ga_\circ^a\,{\lrdrt a}\,\psi$ or $\bar{\psi}M_0\psi$.
The redefinition matrix can be taken either as $1+\delta\hat{U}_I^h$ or $1+\delta\hat{U}_I^h+\delta\hat{U}_V^h$, depending on whether the original operator contains LV coefficients or not.
\eit
The good news is that, we can prove that up to linear order of $h_{\mn}$ and LV coefficients,
there is no need to consider the redefinition induced ``$h$-interaction" arising from the operator
$\mathscr{L}^\mathrm{flat}_\psi=\frac{i}{2}\bar{\psi}\Ga_\circ^a\,{\lrdrt a}\,\psi-\bar{\psi}M_0\psi$
between a pair of one-fermion states $\langle p',\be|\int d^3x\mathscr{L}^\mathrm{flat}_\psi|p,\al\rangle$,
once the Dirac equation is utilized, \ie, the external fermions are on mass-shell.

\section{various eigen-spinors}\label{LVeigenSpinors}
The eigen-spinor in the presence of LV coefficients will be given separately by assuming only one-type LV coefficient
is nonzero. For a more general treatment including non-minimal LV coefficients, the interesting reader can resort to
Ref. \cite{arbitrFermion}\cite{ReisMacro2016}

Firstly, the eigen-spinor for $a$ and $b$ coefficients can be found in \cite{SMEa}, and for completeness, we collect it here. As $a$-term
acts like a shift in 4-momentum, we will give the corresponding eigen-spinor here together with the $b$-term:
\bea\label{eigenSpinor}
u_\al(k)=\left(
           \begin{array}{c}
             \xi^\al \\
             U_{ab}(k)\xi^\al \\
           \end{array}
         \right),
\eea
where $U_{ab}(k)\equiv\frac{[(k^0+a^0)+m-\vec{b}\cdot\vec{\si}][(\vec{k}+\vec{a})\cdot\vec{\si}+b^0]}{[(k^0+a^0)+m]^2-\vec{b}^2}$,
and the two-component spinors $\xi^\al$ satisfy the eigenvalue equation given by (A4-A5) in \cite{SMEa}.
It is clear from the above consideration that $a^\mu$ serve as a pure shift in 4-momentum, and thus is usually ignored
due to field redefinition. However, we keep $a$-term here as we will see gravity concerns the $a$-term.
For calculational convenience, we also note
\bea&&\label{UbM}
U_{b}(k)=(k^0+m+\vec{b}\cdot\vec{\si})^{-1}(\vec{k}\cdot\vec{\si}+b^0)\nn&&~~
\simeq\frac{b^0+\vec{k}\cdot\vec{\si}}{\omega_0+m}-\frac{i\vec{b}\times\vec{k}\cdot\vec{\si}+\vec{b}\cdot\vec{k}+\delta\omega\,\vec{k}\cdot\vec{\si}}{(\omega_0+m)^2},
\eea
where $\omega_0=\sqrt{\vec{k}^2+m^2}$ and $\delta\omega\equiv k^0-\omega_0$.
As for $b,~d,~g,~H$ type LV coefficients, the 4-fold degeneracy between all 4 eigen-spinors is completely broken!
and the explicit form of $k^0$ (and hence $\delta\omega$) is very complicated even at linear order of LV coefficients.
Depends on the nature of LV coefficients, a simple form of $\delta\omega$ maybe obtained.
For example, if $b^2>0$, in an observer frame where $b^0=0$, $\delta\omega=(-1)^\al[m^2\vec{b}^2+(\vec{b}\cdot\vec{k})^2]^{\hf}/\omega_0$,
where $\al=1,2$ and denotes the two spin d.o.f. of $\xi^\al$.

Then we turn to $H$ coefficients. The corresponding operator in the Lagrangian is $-\hf H_{ab}\si^{ab}$, which implies that
$H_{ab}=-H_{ba}$. The antisymmetric property indicates that we can define two vectors, $\vec{H}^i\equiv H_{0i}$ and
$\vec{\tilde{H}}^i\equiv\hf\epsilon_{ijk}H_{jk}$. Then the eigen-spinor $u^\al(k)$ can still be written in the form of (\ref{eigenSpinor}),
only by replacing $U_{ab}(k)$ with
\bea&&\label{UHM}
\hspace{-3mm}U_H(k)=\frac{(k^0+m-\vec{\si}\cdot\vec{\tilde{H}})\,\,\vec{\si}\cdot(\vec{k}-i\vec{H})}{(k^0+m)^2-\vec{\tilde{H}}^2}\nn&&~~
\simeq\frac{\vec{\si}\cdot(\vec{k}-i\vec{H})}{\omega_0+m}-\frac{\vec{k}\cdot\vec{\tilde{H}}+i\vec{\tilde{H}}\times\vec{k}\cdot\vec{\si}
+\delta\omega\,\vec{\si}\cdot\vec{k}}{(\omega_0+m)^2},
\eea
where we also only keep linear order corrections due to LV $H$-coefficients.
Note $\omega_0$ and $\delta\omega$ are also defined as the above, but now $\delta\omega$ only receives LV corrections from $H$-coefficient.

Naively, we can also obtain
\bea\label{UdMi}&&
U_d(k)=[k^0+m-\vec{\c{d}}\cdot\vec{\si}]^{-1}[\c{d}_{0}+\vec{k}\cdot\vec{\si}]\nonumber
\eea
where we defined $\c{d}_\mu\equiv d_{\mn}k^\nu$ for simplicity.
However, as mentioned in the main text that a proper treatment of $c,~d$ terms involves field redefinition,
which gives the correct $U_c(k),~U_d(k)$ by the procedure in getting $U_b(k)$.
The $U_c(k),~U_d(k)$ up to linear order of $c,~d$ coefficients are shown below,
\bea&&\label{UdMc}
\hspace{-5mm}U_d(k)=\left[k^0+m(1+d_{j0}\si^j)-\tilde{d}_{ij}\si^ip^j\right]^{-1}(\vec{\si}+2\vec{d})\cdot\vec{p}\nn&&~~~~
\simeq\frac{\vec{k}\cdot(\vec{\si}+2\vec{d})}{\omega_0+m}-\frac{(md_{j0}-\tilde{d}_{ji}k^i)k^j}{(\omega_0+m)^2}\nn&&~~~~~~
-\frac{i\epsilon_{jkl}(md_{j0}-\tilde{d}_{ji}k^i)k^k\si^l+\delta\omega\,\vec{\si}\cdot\vec{k}}{(\omega_0+m)^2},\\\label{UcMc}&&
\hspace{-5mm}U_c(k)=\frac{\vec{\si}\cdot\vec{p}-\tilde{c}_{ij}\si^ip^j}{k^0+m(1-c_{00})+2\vec{c}\cdot\vec{p}}\nn&&~~~~
\simeq\frac{\vec{k}\cdot\vec{\si}-\tilde{c}_{ij}\si^ik^j}{\omega_0+m}+\frac{mc_{00}-2\vec{c}\cdot\vec{k}-\delta\omega}
{(\omega_0+m)^2}\vec{k}\cdot\vec{\si},
\eea
where it is easy to separate the formally LV contributions from the LI one, $\frac{\vec{k}\cdot\vec{\si}}{\omega_0+m}$.
To obtain the correction $\delta\omega$ for $b,~H,~c,~d$ coefficients, we'd better find out their explicit dispersion relations,
which can be found in \cite{Lehnert2004}\cite{ZXDR}.
They all share the similar form
\bea\label{GDRbdH}
[(k^0)^2-\omega_0^2+Y^2]^2=4Z^2,
\eea
where
\bea&&
Y^2=\left\{\begin{array}{c}
    b^2 \\
    (\vec{\tilde{H}}^2-\vec{H}^2)\\
    d_{ab}k^b\,d^a_{~c}k^c
  \end{array}\right.,\\&&
Z^2=\left\{\begin{array}{c}
    (k\cdot b)^2-k^2b^2 \\
    {H^*}^{\mn}k_\mu{H^*}_{\zeta\nu}k^\zeta-(\frac{1}{4}{H^*}^{\mn}H_{\mn})^2\\
    (k^ad_{ab}k^b)^2-k^2\,d_{ab}k^b\,d^a_{~c}k^c
  \end{array}\right.,
\eea
where ${H^*}^{\mn}\equiv\hf\epsilon^{\mn\al\be}H_{\al\be}$, and the three rows of $Y^2$ and $Z^2$
correspond to $b,~H,~d$ terms, respectively.
While for $c$ term, the dispersion relation is simply $(c_{\mn}+\eta_{\mn})k^\nu(c^{\mu\rho}+\eta^{\mu\rho})k_\rho+m^2=0$,
which is spin-independent, and thus leads to much greater calculational simplicity for $\delta\omega$.
From the exact dispersion relation (\ref{GDRbdH}), we can readily obtain
\bea\label{deltaome}&&
\delta\omega
=\omega_0\left[\sqrt{1\pm\frac{2Z}{\omega_0^2}-\frac{Y^2}{2\omega_0^2}}-1\right]
\simeq\pm\frac{Z}{\omega_0}\simeq\pm\frac{Z}{m},
\eea
where we ignore $Y^2$, as $Y^2,~Z^2\sim\mathcal{O}(X^2)$ are at least of second order of a generic LV coefficient $X$.
Note that the sign ambiguity associated to $\delta\omega$ for $b,~d,~H$ coefficients reflects the fact that
the corresponding terms are spin-dependent.
Thus the degeneracy for dispersion relations between relevant eigen-spinors are completely removed.
However, since in calculating matrix elements, $\delta\omega$ in effect acts on the two-component spinor $\xi^\al$,
see Eq. (\ref{eigenSpinor}-\ref{UHM}), and thus the ambiguity can be removed by some kinds of ``eigen-equations".
These ``eigen-equations" can be obtained by taking the ``square root" of the exact quartic dispersion relation
$\mathrm{det}\left[\Ga\cdot k+M\right]=0$.

Taking the $b$-term as an example. Left multiplying the $k$-space positive Dirac equation
$(\ga\cdot k+m+b\cdot\ga_5\ga)u(k)=0$ by $(m-\ga\cdot k-b\cdot\ga_5\ga)$, we get
\bea&&
\hspace{-5mm}-(\ga\cdot k-m+b\cdot\ga_5\ga)(\ga\cdot k-m+b\cdot\ga_5\ga)u(k)\nn&&
\hspace{-5mm}=(k^2+m^2-b^2+2ib_\mu k_\nu\ga_5\si^{\mn})u(k)\nn&&
\hspace{-5mm}=\left(
   \begin{array}{cc}
     K^2+2k_{b\si} & 2i\vec{b}\times\vec{k}\cdot\vec{\si} \\
     2i\vec{b}\times\vec{k}\cdot\vec{\si} & K^2+2k_{b\si} \\
   \end{array}
 \right)\left(
          \begin{array}{c}
            \xi^\al \\
            U_d(k)\xi^\al \\
          \end{array}
        \right)=0,
\eea
where $K^2\equiv k^2+m^2-b^2$ and $k_{b\si}\equiv(b^0\vec{k}-k^0\vec{b})\cdot\vec{\si}$.
Reserving only $b$ terms to linear order, the upper equation
$\left[k^2+m^2+2(b^0\vec{k}-k^0\vec{b})\cdot\vec{\si}+2i\vec{b}\times\vec{k}\cdot\vec{\si}U_0(k)\right]\xi^\al=0$
for Pauli spinor can be rearranged as
\bea&&
\hspace{-8mm}(k^0+\omega_0)\delta\omega\xi^\al=\left[4b^{[0}k^{j]}\si^j+2i\vec{b}\times\vec{k}
\cdot\vec{\si}U_0(k)\right]\xi^\al,
\eea
which leads to the eigen-equation for LV correction $\delta\omega$
\bea&&
\hspace{-10mm}\delta\omega\xi^\al=\left[-\vec{b}\cdot\vec{\si}+\frac{b^0\vec{k}\cdot\vec{\si}}{\omega_0}
+\frac{\vec{k}^2\vec{b}\cdot\vec{\si}-(\vec{k}\cdot\vec{b})(\vec{k}\cdot\vec{\si})}{\omega_0(\omega_0+m)}\right]\xi^\al,
\eea
Similarly for $H,~d,~c$ terms, we have
\bea&&
\delta\omega\xi^\al=
\left[\vec{\tilde{H}}\cdot\vec{\si}+\frac{\vec{H}\times\vec{k}\cdot\vec{\si}}{\omega_0}
-\frac{\vec{\tilde{H}}\cdot\vec{k}\,\vec{\si}\cdot\vec{k}}{\omega_0(\omega_0+m)}\right]\xi^\al,\\&&
\delta\omega\xi^\al=\left[(md_{j0}+\tilde{d}_{ji}k^i)\si^j+2\frac{\vec{d}\cdot\vec{k}\,\vec{\si}\cdot\vec{k}}{\omega_0}\right.
\nn&&~~
\left.-\frac{(md_{j0}+i\epsilon_{ikl}d_{ij}k^k\si^l)k^j\,\vec{\si}\cdot\vec{k}}{\omega_0(\omega_0+m)}\right]\xi^\al,
\\\label{deltac}&&
\delta\omega=-2c_{(0j)}k^j-c_{(ij)}\frac{k^ik^j}{\omega_0}-c_{00}\omega_0,
\eea
where due to the spin-independence, there is no need to act on Pauli spinors for $c$ coefficient, compared with other LV coefficients,
and we choose the positive sign corresponding to electron's dispersion relation instead of positron's.
Substituting these $\delta\omega$ back into (\ref{UbM},\ref{UHM},\ref{UdMc},\ref{UcMc}), we can get
\bea&&\label{UbkNR}
U_b(k)=\frac{\vec{k}\cdot\vec{\si}+b^0}{\omega_0+m}-\frac{2i\vec{b}\times\vec{k}\cdot\vec{\si}}{(\omega_0+m)^2}+\mathcal{O}(\omega_0^{-3})
\nn&&~~~~
\stackrel{\mathrm{NR}}{\simeq}\frac{\vec{k}\cdot\vec{\si}+b^0}{2m}+\frac{i\vec{k}\times\vec{b}\cdot\vec{\si}}{2m^2},
\eea
\bea\label{UHkNR}&&
U_H(k)=\frac{\vec{\si}\cdot(\vec{k}-i\vec{H})}{\omega_0+m}-\frac{2\vec{k}\cdot\vec{\tilde{H}}}{(\omega_0+m)^2}
+\mathcal{O}(\omega_0^{-3})\nn&&~~~~
\stackrel{\mathrm{NR}}{\simeq}\frac{\vec{\si}\cdot(\vec{k}-i\vec{H})}{2m}-\frac{\vec{k}\cdot\vec{\tilde{H}}}{2m^2},
\eea
\bea\label{UdkNR}&&
U_d(k)=\frac{\vec{k}\cdot(\vec{\si}+2\vec{d})}{\omega_0+m}
-2\frac{(md_{j0}-i\epsilon_{ikl}\tilde{d}_{ij}k^k\si^l)k^j}{(\omega_0+m)^2}+\mathcal{O}(m^{-3})\nn&&~~~~
\stackrel{\mathrm{NR}}{\simeq}\frac{\vec{k}\cdot\vec{\si}+d_{0j}k^j}{2m}+\frac{i\epsilon_{jkl}d_{ji}k^kk^i\si^l}{2m^2},
\eea
\bea\label{UckNR}&&
\hspace{-8mm}
U_c(k)=\frac{\vec{\si}\cdot\vec{k}-c_{ij}\si^ik^j}{\omega_0+m}+\mathcal{O}(\omega_0^{-3})
\stackrel{\mathrm{NR}}{\simeq}\frac{\vec{\si}\cdot\vec{k}-c_{ij}\si^ik^j}{2m},
\eea
where we have ignored the terms suppressed by higher orders than $\omega_0^2$ (or $m^{-2}$).
The $\mathrm{NR}$ at the last step means we adopt the non-relativistic approximation.
Substituting these $U(k)$ in (\ref{UbkNR}-\ref{UckNR}) into
\bea\label{eigenSpinorbdH}
u_\al(k)_X=\left(
           \begin{array}{c}
             \xi^\al \\
             \eta^\al \\
           \end{array}
         \right)_X=\left(
           \begin{array}{c}
             \xi^\al \\
             U_X(k)\xi^\al \\
           \end{array}
         \right),
\eea
where $X$ in the subscript refers to $b,~H,~d,~c$, we obtain the LV corrected positive frequency
eigen-spinors up to linear order of LV coefficients.

\end{document}